\documentclass[sigconf]{acmart}

\AtBeginDocument{%
  \providecommand\BibTeX{{%
    \normalfont B\kern-0.5em{\scshape i\kern-0.25em b}\kern-0.8em\TeX}}}

\copyrightyear{2023}
\acmYear{2023}
\setcopyright{acmlicensed}\acmConference[CHI '23]{Proceedings of the 2023 CHI Conference on Human Factors in Computing Systems}{April 23--28, 2023}{Hamburg, Germany}
\acmBooktitle{Proceedings of the 2023 CHI Conference on Human Factors in Computing Systems (CHI '23), April 23--28, 2023, Hamburg, Germany}
\acmPrice{15.00}
\acmDOI{10.1145/3544548.3581204}
\acmISBN{978-1-4503-9421-5/23/04}

\usepackage{graphicx}
\usepackage{subfig}
\usepackage{cleveref}
\usepackage{amsmath}

\begin{document}

\title{Data, Data, Everywhere: Uncovering Everyday Data Experiences for People with Intellectual and Developmental Disabilities}


\author{Keke Wu}
\affiliation{%
  \institution{University of North Carolina at Chapel Hill}
  \city{Chapel Hill}
  \country{United States}}
\email{kekewu@cs.unc.edu}

\author{Michelle H Tran}
\affiliation{%
  \institution{University of Colorado Boulder}
  \city{Boulder}
  \country{United States}}
\email{michelle.h.tran@colorado.edu}

\author{Emma Petersen}
\affiliation{%
  \institution{University of Colorado Boulder}
  \city{Boulder}
  \country{United States}}
\email{emmajpetersen@gmail.com}

\author{Varsha Koushik}
\affiliation{%
  \institution{Colorado College}
  \city{Colorado Spring}
  \country{United States}}
\email{vkoushik@coloradocollege.edu}

\author{Danielle Albers Szafir}
\affiliation{%
  \institution{University of North Carolina at Chapel Hill}
  \city{Chapel Hill}
  \country{United States}}
\email{danielle.szafir@cs.unc.edu}



\begin{abstract}
Data is everywhere but may not be accessible to everyone. Conventional data visualization tools and guidelines often do not actively consider the specific needs and abilities of people with Intellectual and Developmental Disabilities (IDD), leaving them excluded from data-driven activities and vulnerable to ethical issues. To understand the needs and challenges people with IDD have with data, we conducted 15 semi-structured interviews with individuals with IDD and their caregivers. Our algorithmic interview approach situated data in the lived experiences of people with IDD to uncover otherwise hidden data encounters in their everyday life. Drawing on findings and observations, we characterize how they conceptualize data, when and where they use data, and what barriers exist when they interact with data. We use our results as a lens to reimagine the role of visualization in data accessibility and establish a critical near-term research agenda for cognitively accessible visualization. 
\end{abstract}



\begin{teaserfigure}
\includegraphics[width=\textwidth]{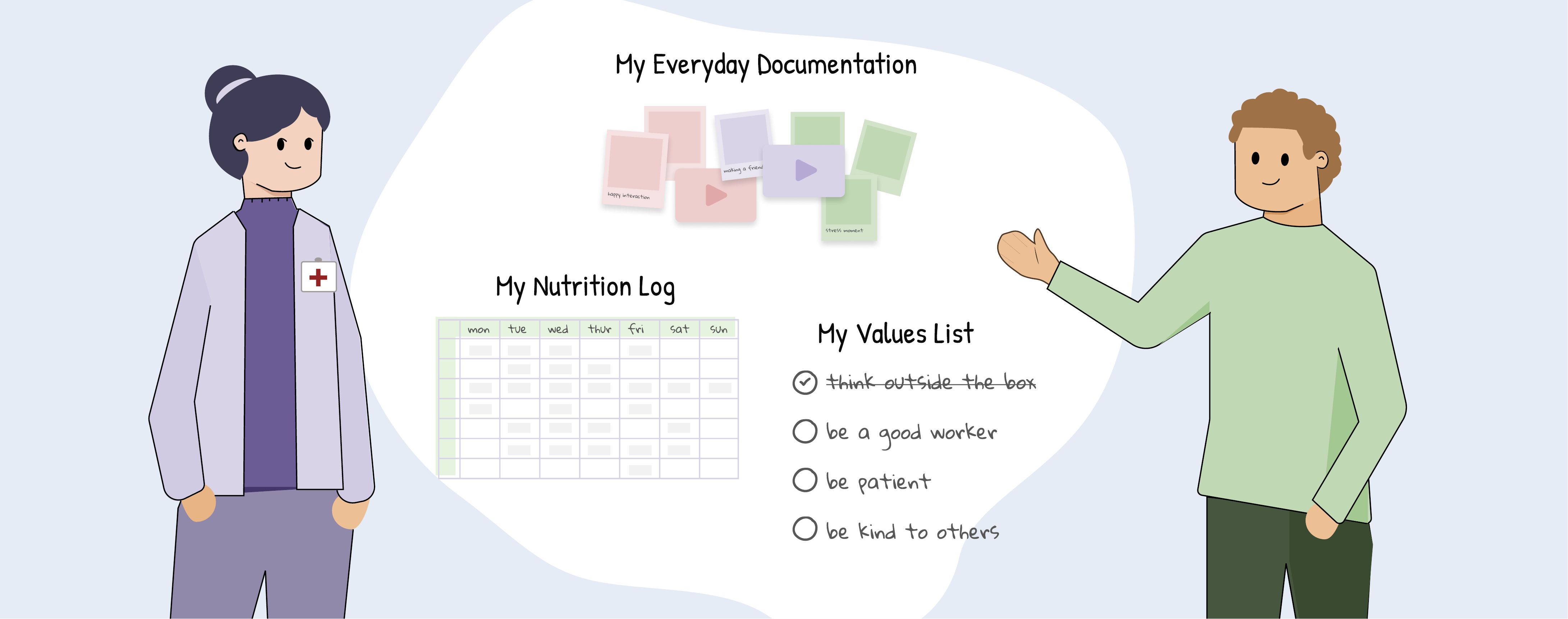}
  \caption{People with Intellectual and Developmental Disabilities (IDD) use personal data to make sense of everyday life. Participants with IDD most frequently used data to track activities of daily living (e.g., nutrition log, checklist, and everyday documentation); however, this data was often invisible to people despite being used in a variety of common scenarios, such as to advocate and shape treatment in healthcare applications. We synthesize limitations in existing approaches to working with data and opportunities for using visualization to enhance data accessibility for people with IDD. For example, participants and their caregivers identified a range of scenarios where data had significant potential to help people with cognitive and communicative impairments better participate in social and self-advocacy situations.}
  \Description{A person is showing his data to a doctor. The data ranges from multimedia files of everyday documentation to nutrition log to a personal values check list.}
  \label{fig:teaser}
\end{teaserfigure}

\maketitle
\section{Introduction}
\label{sec:introduction}
  \vspace*{-0.5 mm}
Data is everywhere. We consciously and unconsciously experience data in nearly all aspects of life, from addressing challenging problems in the workplace to navigating social situations.
Data might be captured as numbers in a spreadsheet, a table within a database, a text file, or a collection of images or videos. Knowing how to work with data empowers people to find solutions to problems, make informed decisions, and develop better understandings of ourselves and the world around us. However, working with data can be complex. Barriers to data access and use can create substantive inequities in society that lead to potentially devastating consequences for self-determination \cite{braddock, shogren_2014_autonomy}, professional autonomy \cite{sarju_2021_nothing}, and even basic healthcare \cite{boyle_2020_the}. 

Visualization aims to make data more useful and accessible by graphically representing complex data relationships and communicating data-driven insights. Data visualizations amplify human cognition to help people more clearly see patterns, trends, and outliers in data \cite{inbook, book}. However, using visualizations requires a viewer to read abstract imagery, estimate statistics, and retain information. These processes require people to employ a combination of cognitive skills, reasoning strategies, and graphical and numerical literacy. These skills usually function differently for people with Intellectual and Developmental Disabilities (e.g., Down Syndrome, Williams Syndrome, Fetal Alcohol Syndrome, Autism, and Cerebral Palsy) \cite{chi}. People with Intellectual and Developmental Disabilities (IDD) typically have significant limitations in intellectual functioning; struggle with learning, reasoning, and problem-solving; and experience developmental delays in language, motor, social, and other cognitive areas. However, these differences are seldom considered in visualization tools and techniques, which are primarily developed for neurotypical consumers. This lack of inclusion has created cognitive barriers for people with IDD to effectively access and use data. In addition, people with IDD also often face challenges in daily behaviors, such as communication, social interaction, and independent living, which may require them to experience data differently from neurotypical individuals.

In this paper, we explore how people with IDD encounter data in everyday life, uncover their perceptions about data, characterize their needs and challenges for data access, and discuss visualization’s role in data accessibility. We conducted 15 semi-structured interviews with people with IDD and their caregivers to investigate the use of data and visualization in everyday contexts. We categorized our findings into three distinct themes summarizing how people with IDD \textit{conceptualize} data, when and where they \textit{use} data, and what challenges exist when they \textit{interact} with data. Drawing on findings and observations, we synthesize the key takeaways for visualization and discuss actionable insights to use visualization to empower people with IDD. Focusing on data use generally instead of low-level tasks or specific visualization types allowed us to ground data in particular decision-making scenarios people with IDD encounter and deeply understand their lived experiences of data. These insights could inspire authentic solutions and novel visualizations to address real-world challenges and provide concrete support in advocacy, social participation, and independent living. 

\textbf{\emph{Contributions:}}
The primary contribution of this paper is an empirically-grounded characterization of how people with IDD encounter data in everyday life with an emphasis on the role of visualizations. Our interviews revealed that participants frequently used personal data for everyday function, self-expression, self-advocacy, and decision-making. However, this data was usually invisible and sometimes inaccessible to people in the community. We discuss actionable takeaways for how visualization can remedy these challenges in design and improve data accessibility for people with IDD. Our exploratory study capturing the lived experience of people with IDD highlights critical opportunities for future data accessibility research and demonstrates the potential for visualization to empower people with IDD. 
\section{Background \& Related Work}
  \vspace*{-0.5mm}
The unique cognitive profiles of people with IDD and challenges introduced by commonly co-occurring disabilities may result in unconventional needs and challenges for working with data.
We ground our study in psychiatry, disability studies, and special education
to define and characterize
Intellectual and Developmental Disabilities (IDD), and then survey accessible ways to communicate information to people with IDD. While few efforts have investigated the intersection of IDD and data, we review progress in 
accessible visualization in general to explore design solutions and practical strategies for 
inclusive information representation. 

\subsection{Intellectual and Developmental Disabilities}
According to the American Association on Intellectual and Developmental Disabilities (AAIDD) \cite{faqs}, Intellectual Disability (ID) is a developmental disability (DD) characterized by significant deficits in intellectual and adaptive functioning of varying severity that present before 22 years of age. ID affects approximately 2\%
of the population, with males being more likely than females to be diagnosed. We use the term IDD, which describes situations in which intellectual disability and other developmental disabilities are present. IDD more generally encompasses a broad spectrum of functioning and health conditions. People with mild IDD can function independently as adults, while people with more severe IDD often need personalized support for housing, occupational, and recreational activities. Depending on its cause, IDD can be stable and non-progressive or worsen over time. After early childhood, the disorder is chronic and usually lasts for an individual's lifetime; however, the severity of the disorder may change with age. Early intervention can improve adaptive skills and appropriate support, such as active caregivers or community programs, may allow a person with IDD to actively participate in society \cite{boat_2015_mental, sturmey_2014_evidencebased}. In our discussions with people with IDD, caregivers, and community advocates, people with IDD increasingly need support working with data for a variety of applications, ranging from personal informatics (e.g., healthcare or budgeting) to public advocacy (e.g., participating in civic discussions around community policies) \cite{braddock, tuffreywijne_2006_people, accessibleinfo}.
                                           
Each individual with IDD has distinct strengths and weaknesses in their abilities but universally faces functional impairments in real-life skills, such as understanding rules, navigating daily living tasks, and participating in family, school, and community activities \cite{diagnostic}. Many neurodevelopmental, psychiatric, and medical disorders co-occur with IDD, especially communication disorders, learning disabilities, epilepsy, and various genetically transmitted conditions \cite{diagnostic}. Studies indicate that at least 25\% of people with IDD may have significant psychiatric problems, 
including 
increased rates of schizophrenia, depression, and ADHD \cite{psychiatric, bouras_holt_2007}. A major factor for these mental health problems stems from communication impairment, which limits a person's ability to express frustration and/or explain underlying physical or emotional distress \cite{marrus_2017_intellectual}. Due to the large number of co-occurring conditions, IDD tends to be complex and uncertain to diagnose. It is frequently treated as a set of disabilities rather than individual diagnoses, and people with IDD usually need multifaceted support in their interactions with data, which may require visualizations to be designed and used differently. In this paper, we characterize everyday data access challenges and opportunities for people with IDD to understand how visualization tools can better accommodate diverse communicative, cognitive, and expressive needs in design. 

\subsection{Information Communication for People with IDD}
People 
typically
work with data to extract or communicate information \cite{inbook}. Accessible information must be physically and cognitively accessible, adapting to different people, their needs, and abilities \cite{information};
however, we have limited guidance for effectively adapting data to diverse abilities.
People with IDD vary in their ability to understand written, spoken, pictorial, numerical or sign language and express themselves through these various media \cite{Hassiotis2013-ge}. Most people with IDD use speech or sign as their predominant form of communication \cite{Hassiotis2013-ge} and primarily receive information verbally \cite{accessibleinfo}. About 60\% people with IDD are able to use symbolic methods such as pictures, symbols, signs, or speech to communicate \cite{learning_disabilities}. Some people lacking verbal abilities may be able to respond to pictorial narratives \cite{boardman_bernal_hollins_2014}. Those who are nonverbal and only use pre-symbolic communication rely largely on people around them, such as peers and caregivers, to anticipate their needs and interpret their vocalizations, facial expressions, and body language \cite{OKane2016-dl}. These differences make creating accessible information displays highly complex as accessible solutions must address diverse individual abilities, communication strategies, and preferences.

Recent resources aim to support better communication by and with people with IDD. People First published a guide to authoring online or printed information in the Easy Read format \cite{easyread}. They suggest using simplified texts and plain language, managing visual complexity, and adding meaning to words using images. Beyond Words uses pictures to create scenarios for people with IDD to discuss health and social care issues with doctors \cite{beyondwords}. These pictorial stories actively support communication and memory by evoking emotional experience and expression. Visual communication is also used to help people overcome limitations in daily function. For example, individuals with IDD usually struggle with time management and often require prompts to finish daily tasks, such as taking a shower, getting dressed, having breakfast, and seeing a doctor. Visual schedules (also known as \emph{sequence strips}) illustrate an individual's daily routine, showing them what should happen next, helping them understand temporal concepts, and reminding them of important activities \cite{anderson}. Step-by-step illustration and pictorial communication of real-world contexts have helped people with IDD to understand everyday activities, live a more independent life, and better handle and navigate social situations. However, these 
static images cover only restricted scenarios and activities. They do not support other key areas of social participation, such as education, employment, and civic engagement. Intelligently designed visualizations can represent complex relationships in multifaceted data and may empower people with IDD to participate in data-driven conversations in all areas of society.

\subsection{Inclusive Design \& Accessible Visualization}
Carefully crafted interactive visualizations have the potential to increase data access for people with IDD, supporting their autonomous function, communication, and understanding of a variety of social situations.
However, traditional visualizations are 
primarily developed for neurotypical audiences. 
Reading them requires a unique combination of cognitive skills, which may hinder people with IDD in using visualizations: they face impairments in many relevant cognitive abilities (e.g., memory, language, attention, and perceptual speed \cite{faqs, diagnostic}); use different strategies in abstract thinking, learning, and reasoning \cite{math, article}; have significantly restricted access to traditional education; and receive limited exposure to statistical and mathematical training at school \cite{education}. All of these factors may inhibit the use of visualization as traditionally characterized and pose significant challenges for people with IDD \cite{chi}, yet we have limited insight into what these challenges are, when they arise, and how to best address them.  

However, data visualization can reach broad audiences \cite{broader} and has the potential to empower people from all walks of life, including those with a range of abilities \cite{diversity, disability, viscomm}, when designed with attention to the needs of the target user. Past work in accessible visualization has illustrated barriers to access \cite{Kim2021AccessibleVD,chi} as well as methods \cite{lundgard2019sociotechnical,Chundury} and techniques \cite{text-model,holloway20193d,yang2020tactile,zhao2008data} to promote data access. 
For example, Kim et al. \cite{Kim2021AccessibleVD} introduced a design space for accessible visualization drawn from 20 years of work on visualization accessibility. 
Chundury et al. \cite{Chundury} interviewed blind orientation and mobility coaches to reveal possible design practices for conveying spatial information using sonification and auralization. 
G\"{o}tzelmann \cite{tactile} presented a concept for 3D printed tactile graphics coupling tactile exploration and auditory feedback to improve the access to information for visually impaired people.
Lundgard et al. \cite{text-model} developed a conceptual model to communicate contextualized insights in visualization using natural language descriptions. 
Elavsky et al. \cite{Chartability} introduced Chartability, a set of heuristics to help visualization practitioners evaluate the accessibility of data-driven visualizations across various environments, platforms, and contexts.

Although visualization accessibility is gaining increasing attention, the primary focus of past research has been on visual accessibility.
However, physical and cognitive abilities can also influence how people work with data visualizations \cite{disability}. For example, past studies have investigated how particular designs may induce seizures in people with epilepsy \cite{conti2005attacking,south}. 
Of particular relevance to this work, Wu et al. \cite{chi} conducted a mixed methods experiment to understand how people with IDD interpret common data visualizations in the context of financial self-advocacy. The experiment revealed that conventional visualization guidelines may inadvertently create barriers to data use and provided insight into the reasoning strategies and visualization preferences of people with IDD. 
However, the study focused on a niche application of visualization and specific data context. We seek to broaden our understanding of cognitive accessibility by investigating the key scenarios and needs people with IDD have for working with data and visualizations. 

This paper applies a human-centered approach to understanding the use of data and visualization among people with IDD in everyday contexts. Our aim is to characterize the data needs and perceptions of people with IDD to identify key problems for accessible data use, 
provide designers with an awareness of critical issues in visualizations, 
and to understand practices that may enhance visualization accessibility. We explore these concepts to help establish a research agenda for cognitively accessible visualization grounded in the authentic lived experiences of people with IDD.
 \section{Methods} 
\label{sec:method}
To understand the most critical barriers to data access for people with IDD, we first need to understand the relationship between people with IDD and data. Following a human-centered approach to gathering the lived experiences of this population, we conducted a series of 15 semi-structured interviews to discover how they encounter data and visualization. Given that abstract thinking and verbal communication are two of the most prominent challenges for people with IDD ~\cite{faqs}, we adopted an ``algorithmic'' interview process that situated data in their everyday life to actively uncover otherwise hidden data while also acknowledging their agency. This interview approach was designed specifically for the present study and was tested and refined during pilot studies with collaborators with IDD. This highly structured interview process, as recommended by our expert collaborators, could help avoid unnecessary confusion or complexities that may disproportionately affect this population.

\subsection{Interview Questions} \label{sec:questions}
Our interviews incorporated twelve questions designed to address three primary areas of interest: (1) individual definitions and perceptions of data, (2) awareness of and scenarios where people encounter data, and (3) current engagement with data and barriers in those interactions. These questions were derived from conversations with subject-matter experts, readings on data and visualization literacy, and discussions within our interdisciplinary team. We piloted our questions in a series of preliminary studies with individuals with IDD who work in leadership roles within various self-advocacy communities to ensure that the range and depth of the questions were sufficient to capture the relevant roles data plays in people's lives and were easily comprehensible given expected communication abilities and educational levels. Our twelve questions, grouped into three themes, were:

\begin{figure*}
	\includegraphics{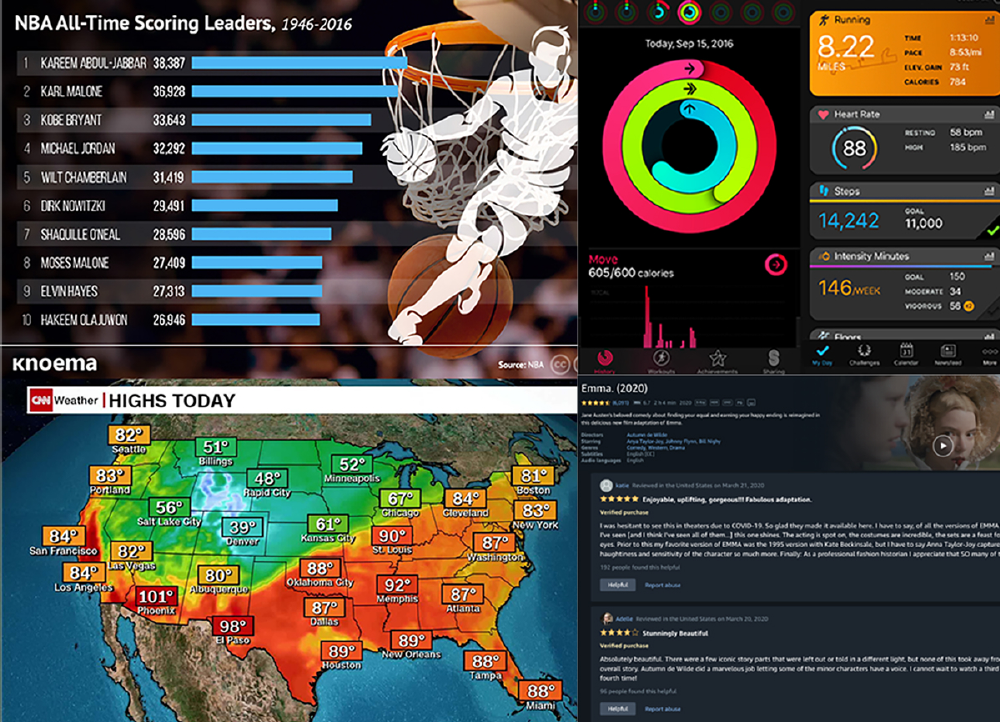}
	\centering
	\caption{Example visualizations used in the interview. }
        \Description {Four example visualizations used in the interview. A bar chart showing NBA ALL-time Scoring Leaders from 1946 to 2016. A weather map of temperature highs in the US. A screenshot of Apple Watch's exercise rings and Garmin technology's personal health panel. A screenshot of the reviews for the movie Emma from Amazon's Prime video.}
	\label{fig:example}
\end{figure*}

\vspace{3pt}
\noindent \textbf{Awareness \& Scenarios:} These questions served as ice breakers to build rapport between the experimenter and the participant as well as to establish baseline communication. The questions probed relevant and important experiences for participants and explored the role that data may have in each individual's major activities as well as other potentially relevant encounters with data. These questions helped establish our expectations about the level of data literacy and disability of each participant and shaped the ensuing conversation. Specific questions included:
\begin{enumerate}
    \item Tell me about yourself. What's your typical day like?
    \item Can you think of any data in the activities you just mentioned? 
    \item If so, what data, in what formats? If not, let us navigate a scenario you just mentioned to see if there is data.
     \item Is there anything else you do regularly to take care of yourself, have fun, express feelings, release stress, etc.?
\end{enumerate}

\vspace{3pt}
\noindent \textbf{Definition \& Perception:} After initial discussions of where data might arise in participants' lives, we explored participants' perceptions of data. We opted to not define data 
up front in order to gather participants' authentic reactions and keep them engaged in the conversation. To encourage personal reflections on data, we asked participants to define data in their own words based on the activities they had previously mentioned. This allowed us to observe participants' comfort talking about data, their conceptions of data, and genuine attitudes towards data. In cases where a participant did not understand the word ``data,'' we worked with their support staff to communicate this intention. Specific questions included:
\begin{enumerate}

    \item Now, let us take a step back, if you were to describe data in your own words, what would that be---what is data?
    \item How do you feel about data? Do you relate to data personally? 
    
\end{enumerate}

\vspace{3pt}
\noindent \textbf{Usage \& Barriers:} To expand on positive and negative experiences with data, we asked a series of questions that explored how and where participants wished to use data. These questions explored current successful and unsuccessful examples of data use, barriers to engagement, and means for exploring data. We additionally walked participants through a series of 22 visualizations to further probe their data perception and awareness, scenarios of data access, and tool use. We asked participants to reflect on the familiarity, utility, and accessibility of the target visualizations (Figure \ref{fig:example}). Specific questions included:
\begin{enumerate}
    \item Why do you feel this way about data? Can you tell me a positive/negative experience?
    \item What are the challenges or barriers you have when you use data?
    \item Do you have any suggestions how we can improve? Do you wish to have any tools or features?
    \item Let us examine these provided examples. Do you find any of them familiar? 
    \item Do you think there is data in there? Has your definition of data changed? 
    \item How do you like the visualizations? What do you read about them? Are they familiar to you? 
    
\end{enumerate}

We used a corpus of visualizations from common platforms for the final part of this section to help reveal how participants' conceptualization of data may have changed over the course of the interview and to understand the perceived accessibility of visualization techniques with concrete examples. Using existing real-world images allowed us to avoid issues around defining and recognizing often invisible visualizations in many areas of interest expressed by participants. The corpus of tested visualizations contained common applications of data in daily life, including isotypes for movie reviews, graphs from common fitness trackers such as the Fitbit or Apple Watch, minimaps in video games, and navigation aids such as Google Maps. In cases where a participant had a preferred app or tool, we shared our screen and navigated a particular example together to understand their decision-making process. The corpus had a total of 22 visualizations across 7 topics, including Music, Shows \& Movies; Travel \& Transportation; Video Games \& Sports; Food, Drink \& Online Shopping; Personal Health \& Finance; News, Information \& Social Media; and Public Policy \& Advocacy. 

We did not ask about low-level details of design (e.g., colors, layout, typography etc.) as that would 
require a certain level of verbal and graphical fluency which many participants with IDD did not have. Our collaborators noted that a lack of fluency combined with questions at that level of detail may have led to alienation or disengagement in
the conversation. Instead, focusing on high-level graphic reading (e.g., ``What do you read from this chart?'' ``Is there something that catches your attention?'' ``Which part you do feel is most overwhelming?'' etc.) gave participants more flexibility, helped us better understand the role of visualization in participants' data interpretation, and offered insights into their sensemaking process. The entire corpus is available at \url{https://osf.io/724ed/?view_only=data-experience}. 

\begin{figure*}
	\includegraphics[width=1\textwidth]{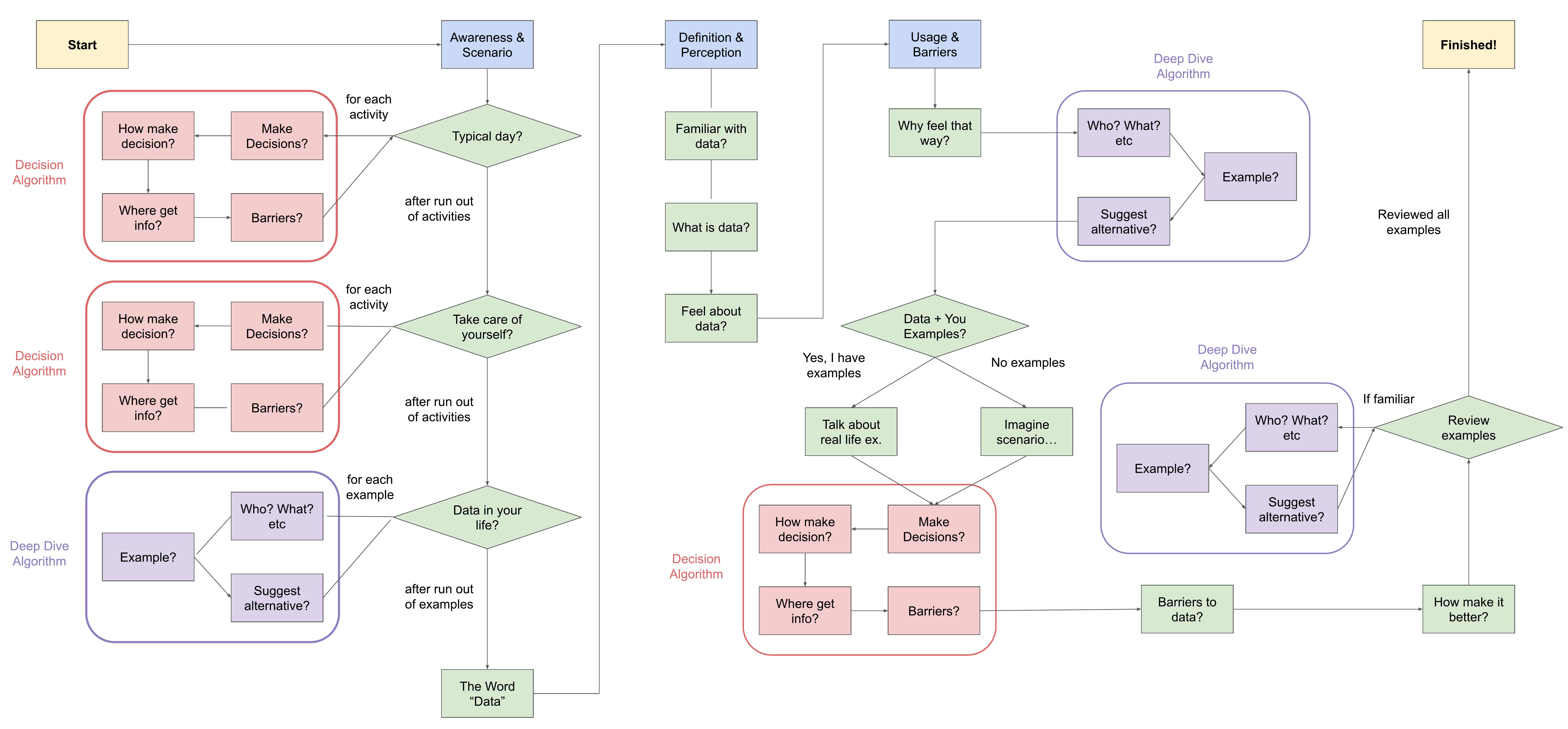}
	\centering
	\caption{We followed an algorithmic process for follow-up questions in our interview studies to minimize limitations of communicative impairments. The general flow for gathering information about personal data experiences followed the process summarized above. For questions pertaining to particular scenarios, we explored how data impacted decision-making using an algorithmic process to avoid complications due to communicative limitations. Blue squares denote the three question categories of interests, green squares/diamonds are specific questions, pink squares are the decision algorithm, and purple squares are the deep dive algorithm.} 
        \Description{A flowchart illustrating the algorithmic interview process. It starts from the top left with asking participants about their awareness and scenarios of data. They are prompted to talk about typical day, and for each activity, the team will use a decision making algorithm to probe their decision-making process and a deep dive algorithm to ask follow-up questions. Then advancing to the definition and perception of data, participants will first give opinions on data, and then be asked about concrete examples of their data experiences, the usage and barriers they have. Finally, participants will look at a data slide deck and comment on the visualizations.}
	\label{fig:flowchart}
\end{figure*}

\subsection{Procedure}
We conducted the interviews over Zoom, with each interview lasting about 1 hour. The conversation started with a brief introduction of the study's objectives. The pre-interview phase was composed of open-ended demographic questions (e.g., age, voluntary disability disclosure, self-rated data literacy, living independently or with support). Given the expected complexity of literacy tests and feedback from pilot participants, we did not use formal 
literacy assessments, but asked participants to self report data literacy in this phase to (1) situate the conversation around data, and (2) understand how self-perceived data literacy may shape attitudes, definitions and comfort discussing data. We then conducted the interview using the sequence of questions outlined in \S \ref{sec:questions}. For questions involving specific examples, we used screenshare to walk through a slideshow of images containing each of the examples in sequence.

Due to limitations in communication associated with many IDDs, we composed a set of pre-scripted follow-up questions based on participants' answers which were developed during our pilot studies. 
These follow-up questions were open-ended but followed a decision tree structure designed to gather specific details about each participant's data experiences and data-based decisions. We adopted this highly structured approach as suggested by our expert collaborators: Given the diversity of audience and the sensitivity of the topic, we needed to be consistent, clear, and intentional in our data collection process to be mindful of the needs and abilities of different participants, proactively avoiding unnecessary cognitive load or stress while maintaining focused engagement. Our data-driven interview flow is illustrated in Figure \ref{fig:flowchart}.

\vspace{6pt}
\noindent \textbf{``Decision'' Algorithm:} 
To uncover unconscious data encounters and understand 
how people with IDD make decisions with data in their daily lives, after asking \emph{Awareness \& Scenarios} Question 2 and Question 4, or identifying a familiar visualization in \emph{Usage \& Barrier} Question 4, we would expand on data relevant to the participant's described activity as follows:  
\begin{enumerate}
	\item Do you make any decisions based on that data? 
	\item How do you make the decision? 
	\item Where do you get the information? 
	\item Are there any challenges getting or understanding it?
\end{enumerate}

\noindent \textbf{``Deep Dive'' Algorithm: } 
Similarly, for each data example or personal experience that participants described, we would elaborate on the use of data in individual activities using the following question structure:
\begin{enumerate}
	\item What do you do with this data? Do you share your data with someone? 
	\item Can you show me an example of visualizations or other ways you work with this data?
	\item Do you have suggestions about alternative uses of this visualization or other representations that might be easier?
\end{enumerate}

These questions were repeated for each activity the participant mentioned. 
After we finish asking these questions, we would prompt for more activities and repeat them until the participant no longer provided any additional activities. Additional unscripted follow-up questions were employed as needed and as appropriate to help clarify participant responses.  

We took several precautions to encourage effective communication with participants and authentic representation of their answers. 
We intentionally focused on open-ended questions to accommodate unexpected directions or unconventional answers.
Additionally, we sought to minimize the amount of time we spent talking relative to the participant despite potential communication barriers introduced by differences in abilities by leveraging pre-scripted follow-up questions when possible. 
In cases where participants had more severe IDD, we would slightly shift our interview structure, and have them navigate hypothetical scenarios based on their pre-mentioned activities using the "Decision Algorithm" to identify hidden data encounters. The participant's caregiver would often provide guidance or even answer certain questions for the participant. In these cases, we observed the interactions and communication between people with IDD and their caregivers, with special attention to agency issues. 
We requested that all participants had their cameras on during the interview to ensure active participation and to observe patterns in response dynamics between participants and caregivers.

\begin{table*}
	\begin{center}
        \scalebox{0.86}{
		\begin{tabular}{ c|c|c|c|c|c}
			\textbf{Participant} & \textbf{Gender} & \textbf{Age} & \textbf{Data Literacy} & \textbf{Data Confidence} &\textbf{IDD Type} \\
			\hline
			1 & M & 33 & High & High & Autism and ADHD \\ 
			2 & F & 40 & Middle & High & ID  \\ 
			3 & M & 30 & Low & Low & ID \\ 
			4 & M & 22 & Low & Middle & Cerebral Palsy  \\ 
			5 & M & 36 & Low & Middle & Autism and Legally Blind \\ 
			6 & F & 31 & Middle & Low & Autism \\ 
			7 & M & 20 & Low & N/A & Nonverbal Autism  \\ 
			8 & F & 50 & High & High & Fetal Alcohol Syndrome and Borderline Personality Disorder \\ 
			9 & F & 48 & Low & Low & Williams Syndrome  \\ 
			10 & M & 25 & Middle & Low &  ID  \\ 
			11 &  M & 33 & High & High & ID  \\ 
			12 & M & 29 & High & Middle & Autism  \\ 
			13 &  M & 35 & High & Middle & Autism  \\ 
			14 &  F & 23 & Low & Low & Developmental Delay and Sensory Processing Disorder  \\ 
			15 & M & 43 & Middle & High & Autism and Obsessive Compulsive Disorder  \\ 
			\hline
		\end{tabular}}
		\vspace{5pt}
          \Description {A table of our 15 participants' self-reported disabilities and data literacy, as well as our observed data confidence level.}
		\caption{\label{participants} Participants' self-reported disabilities, data literacy, and observed data confidence.}
	\end{center}
\end{table*}

\subsection{Participants} 
Participants were recruited from our local network and other national or regional IDD organizations in the United States. We received approval from our university’s institutional review board before contacting any organizations or individuals. Consent of capable adults and assent of incapable adults' legal guardians were gathered through DocuSign. We recruited 15 participants (5 females and 10 males, 20 to 50 years old) who self-identified as having mild to moderate IDD. The participants were compensated with a \$20 Amazon gift card after finishing the interview. 

IDD encompasses a broad set of diagnoses and severity. 
As specific diagnoses are often difficult to determine and co-morbid with other conditions~\cite{carulla2011intellectual}, we asked participants to voluntarily self-disclose their specific IDD.
Despite a relatively small sample size, our study covered a wide range of demographics and captured rich details of data experiences in everyday contexts. The 15 participants were recruited from 9 different states in the US, covering various severity and types of IDDs and comorbidities, including but not limited to Intellectual Disability, verbal and nonverbal Autism, Cerebral Palsy, Fetal Alcohol Syndrome, Williams Syndrome, Developmental Delay, and Sensory Processing Disorder. Participants had different levels of independence, education, and social participation. 10 participants independently participated in our study, and 5 had caregiver support. Participants had mixed educational backgrounds and generally limited literacy. 11 participants had high school or lower degrees, and 4 of them received college education. 10 participants had a full-time or part-time jobs. 2 also took leadership roles in their local community.
Our participants' self-identified disability profiles are summarized in Table \ref{participants}.

\subsection{Analysis}
All interviews were recorded and transcribed. Two researchers independently coded the responses using a spreadsheet with respect to three topics of interest: (1) definition \& perception; (2) awareness \& scenarios; (3) usage \& barriers. A third coder provided additional feedback based on overlaps and conflicts. We then resolved any conflicting codes and formulated key themes using affinity diagrams and empathy maps, with special attention to common patterns and responses reflecting critical individual differences. These methods were chosen as they are common ways of constructing themes from coded data. Specifically, affinity diagramming ~\cite{pernice_2018} allows multiple team members to collaboratively sort out patterns emerged from the study, and empathy mapping ~\cite{gibbons_2018} helps the team understand the reason behind some actions a person takes, which is especially critical for understanding the experiences of a diverse and historically marginalized population. Our original coding sheets are available at \url{https://osf.io/724ed/?view_only=data-experience}.

 \section{Findings}
 \label{sec:finding}
We centered our analysis around three distinct themes: (1) how people with IDD conceptualize data; (2) when and where they use data; and (3) what barriers exist when they interact with data. We summarize common patterns, highlight specific examples of these patterns, 
and synthesize core takeaways for visualization. While many of our observations may generalize beyond people with IDD, our emphasis is on the lived experiences of people with IDD, including the range in their abilities and needs.

 \subsection{\textbf{What Is Data? }}

\subsubsection{Stereotypes lead to personal insecurities \& apathy about data.}  \label{sec:apathy}

\vspace{2pt}\noindent
More than half of the participants struggled to accurately define data, and had never thought about its meaning before. Either independently or with the help of support staff, they defined data as information or numbers and thought of it as useful and important to make sure things are accurate. Furthermore, five participants believed data to be a digital document gathered and stored on a hard drive and somewhat related to computer technology. This stereotypical understanding of data sometimes translated to personal insecurity and a sense of apathy about data. 
P9 assumed data was only used by avid technology users, which conflicted with her personal identity with respect to technology: \emph{``Oh boy, I really don't know because I’m not a computer technology kind of gal.''} --- P9 (low literacy, low confidence)

Different levels of data literacy, past data experiences, perceived relevance, and self-confidence in data led people to conceptualize data in different ways. On a scale of 1--10, five participants self-identified as having high data literacy (>6) while ten of them rated middle to low (1-6). We additionally noted each participant's data confidence based on observations of their reactions to questions and comfort describing data experiences. We found that people generally fell into two groups: data-aware individuals (7 participants) who recognized some instances of data in their lives and the potential to use that data, and data-apathetic individuals (8 participants) who were either unaware of any data use or felt data was outside of their control and/or abilities to work with. Data-aware participants 
typically had middle to high self-reported data literacy or confidence with data. These participants defined data in the context of actions, such as advocating for yourself, informing other people, and telling stories. In these cases, data was described as files, spreadsheets, PowerPoint presentations, surveys, pictures and photos, etc. For example, P1, a journalism major, described how they use visual data to tell stories and promote events,  \ \

\emph{``Gathering facts is a form of data that I deal with on a daily basis. I take photographs. I like to use my camera to tell visual stories through data, through visual data, essentially. I shoot photos for the athletics department and those photos are used in a way to promote events.''} --- P1 (ASD \& ADHD, high literacy, high confidence) \ \ 

\vspace{2pt}
P8 defined data in the context of data-driven narratives. They often use multimedia to communicate the potential of Special Olympics athletes, \ \

\vspace{2pt}
\emph{``The words we put on the screen, the videos of us performing on TikTok, the PowerPoints... We use the pictures as data to kind of help reinforce what we're trying to talk about. We're trying to introduce people to the fact that I can do anything you can do.''}  --- P8 (Fetal Alcohol Syndrome \& BPD, high literacy, high confidence) 

\vspace{2pt}
These individuals also viewed data as a powerful and unavoidable social currency,
which left them sometimes powerless and with no
agency over the role of data in their lives. 
P12 talked about how data is everywhere and can be everything, 

\vspace{2pt}
\emph{``There is data at all times just never stops, never sleeps. Data in art is colors, shapes, and details, data in computers is electronic information and code.''} --- P12 (ASD, high literacy, middle confidence) 

\vspace{2pt}
On the other hand, data-apathetic individuals self-identified as having low literacy and/or self-confidence with data, had difficulty defining data, and viewed it as an artifact (e.g., information, facts, etc.) without the context of how it is used. This lack of understanding about data led them to struggle to pinpoint data in everyday life and unable to relate to data on a personal level. For example, P6, a graphic designer, associated data with a CD or hard drive, \ \ 

\vspace{2pt}
\emph{``I know [data] has to do with storing things away, like digitally through the customer's folders. Even if it is not referring to technology because of how data is stored on a CD or a hard drive, I don't know how I can relate that personally to me in this case. I can't give out a clear answer to that.''} --- P6 (ASD, middle literacy, low confidence) 

\vspace{2pt}
To data-apathetic people, data was considered abstract and authoritative, usually accumulated and curated by “people who do studies (e.g., researchers and scientists)” --- P10 (ID, middle literacy, low confidence) --- rather than by themselves as the subjects of the data. 
This perceived lack of agency led people to not want to access data and to hesitate to question data. They were often unaware and expressed disinterest in the influences data has on their own lives and therefore voluntarily deprived themselves of the opportunities to work with data or maintain control over their own data. 

\begin{figure*}
\centering
\subfloat[Data-Enthusiastic (P15)]{\label{fig,a}\includegraphics[width=0.45\linewidth]{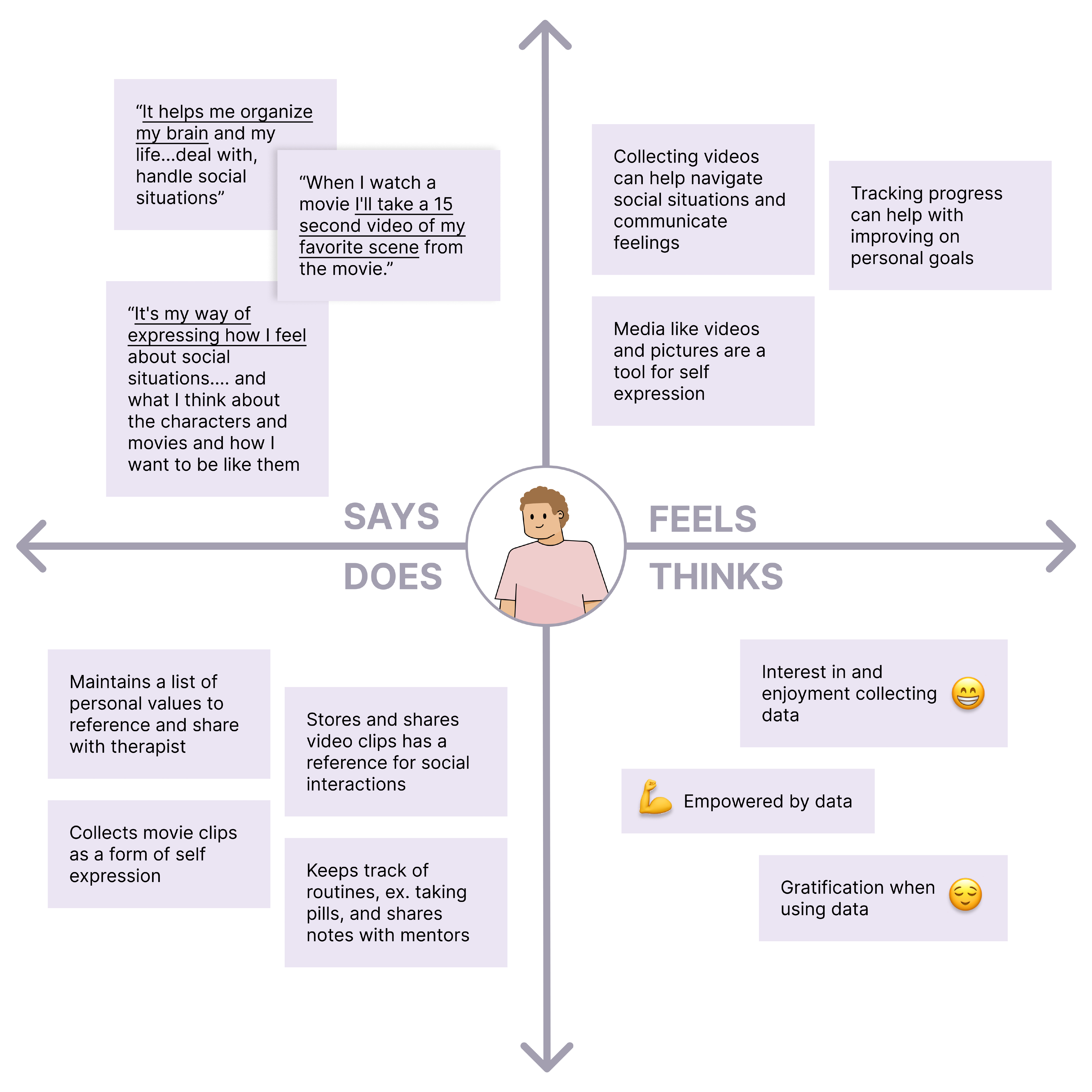}}\qquad
\subfloat[Data-Apathetic (P9)]{\label{fig,b}\includegraphics[width=0.45\linewidth]{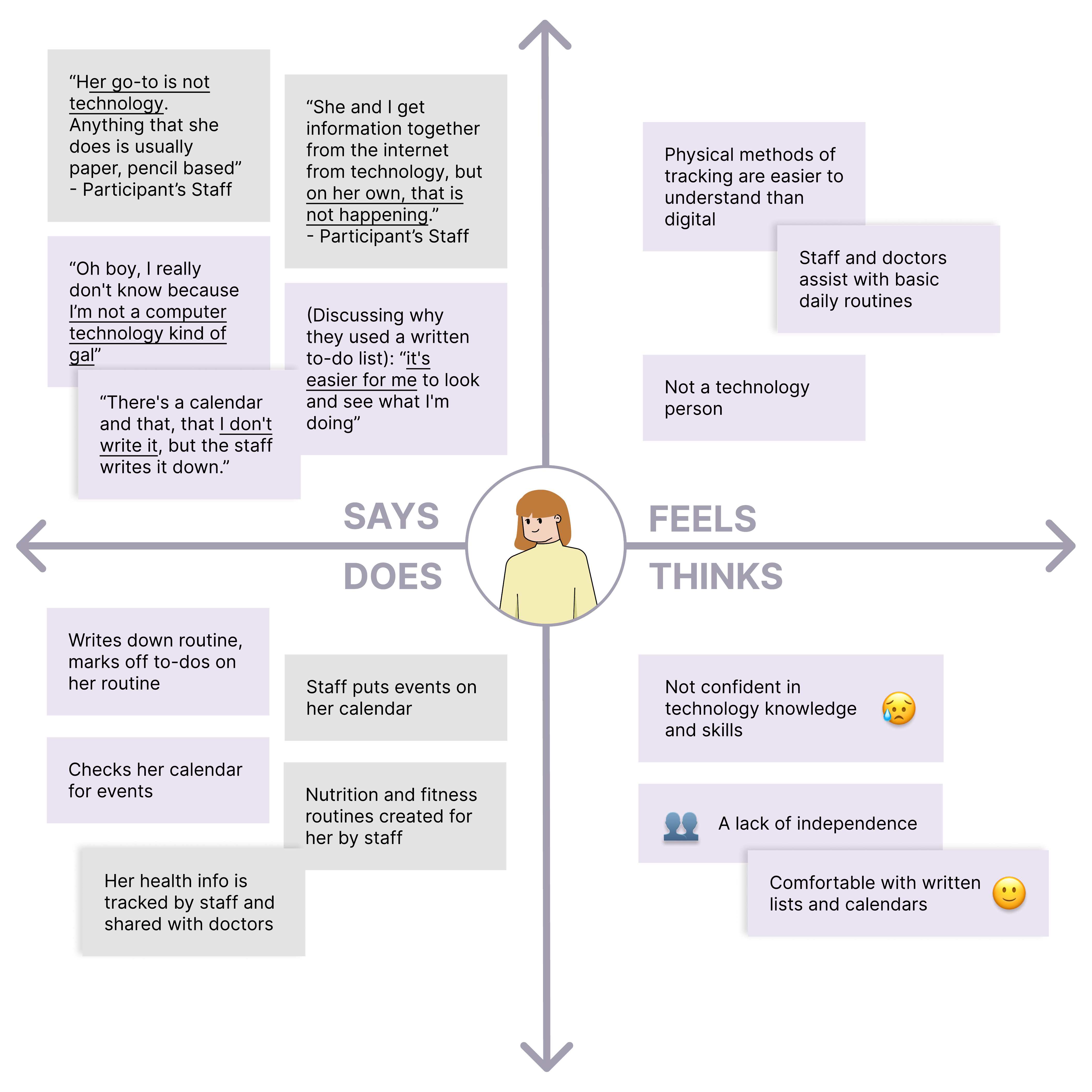}}\\
\caption{Empathy maps that contrast how people with IDD felt and reacted differently to data. Data-enthusiastic participants were more active in the conversation, they gave concrete examples of how they used data and emotionally felt about data; data-apathetic participants associated data with technology and were emotionally detached from data even about themselves. They showed a general lack of interest and confidence in working with data. Caregivers (gray) answered many of their questions.}
\Description{Two quadrants that contrast how different participants say and feel about data, what they think and actually do with data. Data-enthusiastic participants are more positive, using data for personal meaning-making, and are actually managing their own data. Data-apathetic participants are generally negative, not actually aware and using data, and mostly relied on those around them to manage their personal data, even everyday life.}
\label{fig:empathy}
\end{figure*}

To contrast the different levels of awareness, understanding, and emotional connection people had with data, we created two empathy maps in Fig. ~\ref{fig:empathy} to illustrate how a participant's data experience manifests in their thoughts, feelings, words and actions. Data-enthusiastic participants, such as P15 (ASD \& OCD, middle literacy, high confidence), were more active in the conversation, they gave concrete examples of how they used data and emotionally felt about data; data-apathetic participants, such as P9 (Williams Syndrome, low literacy, low confidence), associated data with technology and were emotionally detached from data even about themselves. They showed a general lack of interest and confidence in working with data and relied on their caregivers to answer many of the questions. 

\subsubsection{Predisposition shapes perceptions and perspectives on data.} \ \ \label{sec:perception}

\vspace{2pt}\noindent
Our interviews revealed that participants’ perceptions of data were closely related to how they had personally experienced data, 
their interests and concerns, 
and whether they had a personal goal tied to the data regardless of their disability and data literacy. Personal experiences and individual predispositions shaped not only how people with IDD approach different data, but also how they understand the same data. For example, P2 
positioned their approaches to data in an analogy of
how the grocery manager and customers care about different aspects of data, \ \

\vspace{2pt}
\emph{``If you were owning a grocery store and making the purchases of different products. So then, you know, okay, these are the types of foods people are mostly buying. And people are also using data as a customer. You are looking more at the prices and being like, all right, it's going to be a lot cheaper to buy Jif peanut butter right now than it would be to buy Skippy or whatever.''}  --- P2 (ID, middle literacy, high confidence)

\vspace{2pt}
P11 commented on how personal goals can result in different levels of motivation and awareness in gaming data,

\vspace{2pt} 
\emph{"People don't know that there is data because they want to play the game by itself, not with that analytical mind, but some people love that data by crunching the numbers and doing that stuff. So they can be really good at the game and beat the game."} --- P11 (ID, high literacy, high confidence)

\vspace{2pt} 
Perceptions of the data also changed as a function of data representation and participants' inherent anticipation of that representation. For instance, P13, 
who had autism and social anxiety, explained how paying attention to the content of the documentaries can make data less overwhelming,

\vspace{2pt} 
\emph{"Data can be very overwhelming. That's from my Autism. It has caused a great deal of social anxiety and I try not to really go out at all. When watching films, there is data, like how long the documentaries are, what type of documentary, what genre. But that's not overwhelming, because I'm mostly paying attention to the content."}  --- P13 (ASD, high literacy, middle confidence) \ \ 

\vspace{2pt} 
These examples suggested that perceptions and perspectives of data were inherently subjective and were primarily shaped by personal goals, interests, and attention. People became more aware of data when they involved active attention and sensemaking and when they used data for problem solving. Meanwhile, the platform in which the data appeared helped people with IDD anticipate 
the amount of mental effort and level of difficulty
they would face when working with data in critical applications. People were less aware and less overwhelmed by data that was bundled into activities and systems that helped them better achieve their goals, such as playing well in video games or determining what programs to watch. 

 \subsection{\textbf{How Do You Use Data?}} 
 
\subsubsection{Data is a tool for storytelling.}  \ \  \label{sec:storytellig}

\vspace{2pt}\noindent
Most of our participants 
actively sought opportunities to participate in society. However, the negative narratives and stigmas around people with IDD have created a significant barrier to their participation in many critical facets of daily life. For example, they struggle to get appropriate medical treatments (P2), face employment discrimination (P5), have limited freedom in decision-making (P3), and are often excluded from 
education (P7). One participant, P2, collected data on these problems to find a solution. They conducted a survey to understand disabled individuals' healthcare experience. Analyzing their data helped them to understand and formalize on-going challenges for people with IDD in obtaining medical care,

\vspace{2pt} 
\emph{"The pattern we found was that doctors don't always want to treat patients with disabilities. Some doctors just refuse to see people with disabilities and they kind of link everything of this health challenge to everything is your disability. One thing we have found is that in med school, they don't even discuss how to treat people with intellectual disabilities. So then they don't know, and people are afraid of what they don't know."}  --- P2 (ID, middle literacy, high confidence)

\vspace{2pt}
Another participant envisioned using data as a tool for education, such that relevant stakeholders will be better informed about people with IDD and change their attitudes and perceptions towards people with IDD to address harmful stereotypes. These applications aim to give voice to people with IDD. P5 talked about one such case in the context of employment,

\vspace{2pt} 
\emph{"A lot of employers look at us like we're not important. We get to the interview phase and they find we have this disability and they almost just flat cut you off because we have autism. They’re afraid we will not be able to understand what they're saying, afraid that we will not be able to do the job. But if we have the data that shows what we're able to do and they visually see us do it. That may change their mind."} --- P5 (ASD \& Legally Blind, low literacy, middle confidence)

\vspace{2pt} 
Some participants sought to use data 
for authoritative arguments, such as demonstrating their independence and advocating for a policy change. For example, P4's caregiver pictured using data (i.e., logs of daily activities and accomplishments) to allow people without traditional verbal communication to advocate for themselves in more formal settings, like in a courtroom, 

\vspace{2pt} 
\emph{"I think he's capable of doing a lot more than what he's doing. So we are trying to find different places where he can be exposed to things. We are going to court. We are fighting other administrators' offices to PA (Publication Administrator) so that he can make his own choices. [Data] would give him more supportive decision-making."}  ---on behalf of P4 (Cerebral Palsy, low literacy, middle confidence)

\vspace{2pt} 
Similarly, P7's caregiver described how data (i.e., metrics and recordings of her son's developmental progress over time) could help communicate his abilities and overcome false assertions, 

\vspace{2pt} 
\emph{"I would love to have recorded his progress. The doctor did tell me, `oh, he's never going to talk and he's never going to communicate with you.' So I guess for me, I set out to prove that that was not true. I wish I had that doctor's name. 'Cause I would have sent him this video of my son singing the song. Because from where he was and what they say, he has grown tremendously."}  ---on behalf of P7 (Nonverbal ASD, low literacy, N/A confidence)

\vspace{2pt} 
These results illustrate ways in which data, when carefully compiled and presented, can serve as a persuasive tool to shift stigmas and build powerful narratives for and about people with IDD. Specifically, it can inform others about the needs and abilities of people with IDD, communicate their thinking, and empower them to participate equally in all areas of society. Our interviews specifically point to narrative visualizations as critical for these objectives. 
However, our participants noted two key points for narrative visualizations in these contexts. First, data stories need to shift focus from the data to the people that the data represents, actively communicating the relationships between multiple stakeholders, the challenges they are facing, and their similarities and differences in thinking and behaving in order to proactively identify outliers (e.g., bias, prejudice, and missing links in mutual understanding). Second, visualization can 
link data to aspects of personal identity and help people with IDD foster a positive self-image. Forming this link will require translating their personal goals, missions, and values into relatable visual elements such as shapes, colors, and personally meaningful imagery, and symbolize their sense of self in the discovery of data and making of visualization.

\subsubsection{Data is a medium for expression \& reflection.} \ \ \label{sec:expression}

\vspace{2pt}\noindent
Due to impaired communication skills, many participants faced challenges in interpersonal relationships and experienced significant social anxiety. Many of them developed psychiatric disorders, such as depression (P8), ADHD (P1), and OCD (P15). These individuals typically struggled to empathize with others yet had a desire to engage in social interactions. One participant, P13, collected observations of 
their own behavior, creating datasets of their actions over time and in different contexts to help them reflect on their own behaviors and strategies for social engagement, \emph{"I'm writing all these things down about myself. Mostly just the list of all the issues that I have in social interactions and I keep adding it as I figure things out. I use it to tell the doctor what I need to be tested for and treated for."}  --- P13 (ASD, high literacy, middle confidence)

\vspace{2pt} 
P1
created similar logs on other people's behavior to better understand social interactions, \emph{"I observe human interactions and take notes on my phone. Just sit back and shut up, let people do their thing. And that has helped me so much in understanding human behavior. I use the data to help me understand people and relate to people."}  --- P1 (ASD \& ADHD, high literacy, high confidence)

\vspace{2pt} 
Several participants used sketches, photos, and curated movie scenes to express their thoughts, feelings, and emotional responses to social situations. These artistic media and true-to-life data representations illustrate human interactions with rich details and offer problem-solving strategies on a variety of topics through relatable scenarios and thought-provoking narratives. 
The resulting experiences evoke self-reflection and often leave people with a lasting and pleasant impression, using data for enjoyment. For example, P12 talked about how drawing helped with their creative expression, 

\vspace{2pt}
\emph{"I enjoy drawing a variety of things, like nature, people and objects. I observe what's been going on in my life and at that exact moment. And just taking that from there and applying that to my drawing to get creative in my own way."}  --- P12 (ASD, high literacy, middle confidence)

\vspace{2pt} 
P15 mentioned how 
creating a similar log using 
movie clips helped them express and learn about social interactions,

\vspace{2pt} 
\emph{"When I watch a movie I'll take a 15 second video of my favorite scene from the movie. I share the pictures and videos with my psychiatrist and we talk about how I would handle the situation. It's my way of expressing how I feel about social situations. It helps me organize my brain and my life in dealing with social situations and figure out what the best course of action would be."}  --- P15 (ASD \& OCD, middle literacy, high confidence)

\vspace{2pt} 
In these examples, data served as a reflective device as well as means for expression. People with IDD use data to reflect on themselves, their relationships, and their interactions with society. Real-life data representations, such as photos and movies, can function as a rhetorical medium to facilitate social understanding, self-expression, and release the emotional distress introduced by their communication impairments. 
However, working with these data is challenging. Although our participants noted several systems they have developed to curate and share this data, these systems require substantive effort to manually collect, transcribe, organize, and communicate relevant text or image data. Analytics techniques for working with such data have the potential to help people with IDD use data more efficiently and effectively in these contexts. 
 
\begin{figure*}
\centering
\subfloat[Persona A, 28 years old, has Autism \& Type-1 Diabetes]{\label{fig,a}\includegraphics[width=0.45\linewidth]{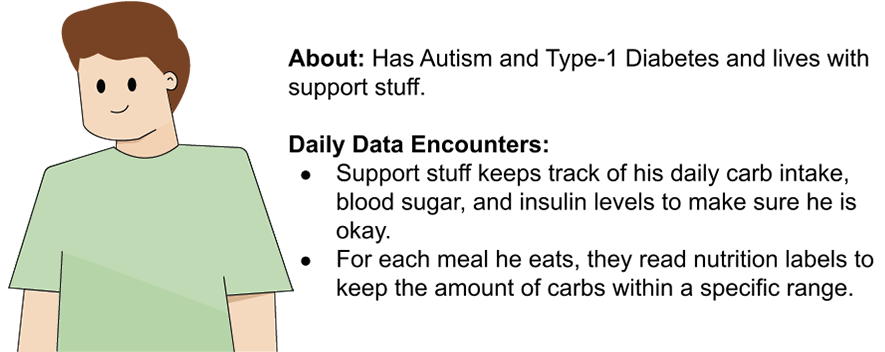}}\qquad
\subfloat[Persona B, 49 years old, has Down Syndrome]{\label{fig,b}\includegraphics[width=0.45\linewidth]{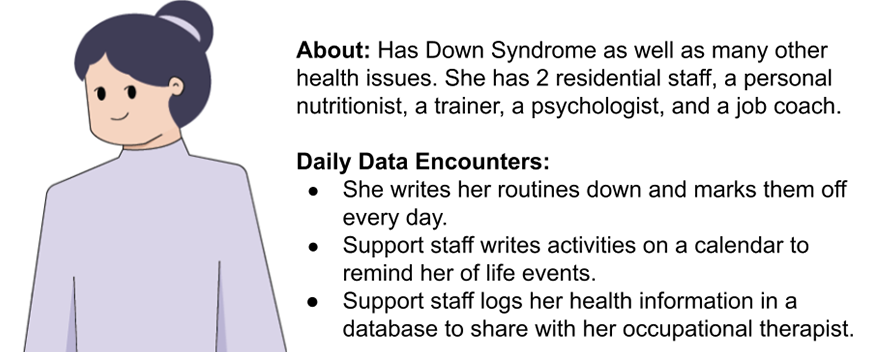}}\\
\caption{Personas that represent two less independent IDD participants and their passive data experiences. Both of them had significant health conditions and were constantly being tracked and monitored by support staff. Persona A wasn't aware of his data and its importance and relied on his caregiver to make sense of it. Persona B's data was collected and shared among multiple professionals, but she was not interested in participating in any of this data-related discussion related to her own well-being.}
\Description{Two personas that represent less independent IDD participants and their passive data experiences.}
\label{fig:persona}
\end{figure*}

\subsubsection{Data is a measure of everyday life.}  \ \  \label{sec:measure}

\vspace{2pt}\noindent
Data was pervasive in participants' lives in many of the same ways as neurotypical populations. For example, participants checked the weather to decide if it was good to go out, what clothes to wear, or even just out of curiosity. Most of them also used interactive maps, read movie ratings and reviews, and browsed websites or social media to keep up with the state of the world. 

Participants experienced visualizations as often as data. When we walked them through the slideshow, most participants could easily recognize at least one of the common data scenarios and visualizations (e.g., bar charts to show sleep hours on a smart watch, movie rating isotype, etc.). However, those visualizations were so familiar and often too embedded in the technology that many participants did not even realize that they were looking at data and not actively taking away meaning from it. In essence, the visualizations were identified as part of a broader technology rather than an interface into data. This raises two opportunities. First, these everyday visualizations displayed on common technology platforms could be a pathway for teaching data literacy and improving data awareness. People with IDD encounter them in the course of daily living, even if they don't recognize these encounters as working with data \textit{per se}. Second, 
these visualizations could benefit from alternative presentations or interactions designed to better capture attention and encourage engagement to promote better sensemaking and active reflection on data.

Due to limited exposure and training in statistics and mathematics, most participants struggled with reading these visualizations
and understanding the information being encoded. They treated the visualizations as decorative interfaces instead of tools for sensemaking. For those who did have some experience with data visualization, they noted several accessibility issues: many of the texts were ``not written in plain language'' (P1); the color choice and layout were often ``way too busy and overstimulating'' (P11); and they felt that ``some visualizations may not work for people with different types of disability (e.g. blind, motor-sensory issues, etc.)'' (P8). The cognitive challenges and inaccessible designs therefore led people to not actively use these visualizations very often in daily life.

More independent participants actively used data to manage daily activities, improve personal behavior, make informed decisions, and take care of themselves and others. For example, P2 talked about how they used a Garmin wearable device to monitor personal health and 
relied on the device's visualizations to make sense of that data, 

\vspace{2pt}
\emph{"I've used it like the steps, the heart rate. It monitors your sleep, how many steps you've taken and things like that. I like the bars here (Fig. \ref{fig:example}). It tells you that one day you had a lot of steps and another day that you barely got any. So if it's tracking days of the week, like, Okay, I know Sundays, I'm not as active as a weekday. It's helpful to know so that you can continue to try to make the right choices."} --- P2 (ID, middle literacy, high confidence)

\vspace{2pt} 
P1 used the summary statistics on Apple Music to better understand their listening habits, 

\vspace{2pt}
\emph{"I love it. It can tell me how long I've been using the app each day, each week, each hour, down to the very minute.
And how many songs I went through the day. And if I'm using it too much, I might just back off and do something else, like take a walk or just taking a nap."}  --- P1 (ASD \& ADHD, high literacy, high confidence)

\vspace{2pt} 
P13 had been on a fitness journey and focused on eating the right types and portions of food. They visually portion controlled each meal using physical measures to track their nutrient intake, 

\vspace{2pt}
\emph{"This is my measuring cup and I have to put one scoop of something that I eat, whether it's beans or veggies or smashed potatoes or green peas... And I lost 20 pounds and I haven't had junk food for two years."}  --- P13 (Autism, high literacy, middle confidence)

\vspace{2pt}
P6 kept track of data to better understand their cat and monitor her health conditions, 

\vspace{2pt}
\emph{"When it comes to knowing about Marble's stuff or knowing how to experiment and what's on her mind or how she feels, I would have to write that all down, like the type of sounds she makes, the time she was vomiting, her reactions to different types of food..."}  --- P6 (ASD, middle literacy, low confidence)

\vspace{2pt}
While data is everywhere, it may not be visible and accessible to everyone. For those less independent participants, their experiences with data were mostly passive, being constantly tracked and measured by caregivers and shared among multiple professionals to make decisions on their behalf without their awareness. As many of the described scenarios were deeply personal and often disclosed private information, we constructed two personas \cite{harley_2015} in Fig. \ref{fig:persona} assembled from specific conditions participants described to represent their data encounters in everyday life.

Persona A has type-1 diabetes in addition to Autism and lives with support staff. The staff keeps track of his daily carbohydrate intake, blood sugar, and insulin levels to monitor his condition and adjust his daily regime in response to this data. For each type of food and each meal he has, they read nutrition labels and keep the amount of carbohydrates within a certain range. However, the staff lacks effective means for sharing this data with Persona A, limiting his ability to participate in his medical care, understand decisions made for him, and decreasing his personal privacy with respect to his food consumption and other relevant factors of his daily activities (e.g., activity levels).

Persona B has Down Syndrome and other health issues and needs support staff at all hours. She works with 
residential staff, a personal nutritionist, a trainer, a psychologist, and a job coach. Persona B writes her routines (e.g., take a shower, brush teeth, take medicine) on a piece of paper and marks them off every day before going to bed, using a similar metaphor to sequence strips \cite{anderson}. Additionally, a support staff member writes things on a calendar to remind her of important events and logs her detailed health information (e.g., exhibited mood, time in activities, dietary intake, etc.) everyday to share with Persona B’s occupational therapist. Information is exchanged between professionals about Persona B's standards of care, but Persona B only has limited access to this data despite her ability to actively communicate with doctors and support staff. 

In all of these cases, data is collected as a means of mediating interactions between a person with IDD and their caregivers. However, these cases also generally 
exclude the individual with IDD 
from the decision-making process. Instead, data is collected about them without means for transparency or collaboration. In some cases, data may be tracked in parallel (e.g., Persona B). However, in all cases, data is hidden from the person the data describes and that individual has little to no agency over its use. 

Visualization can serve as a mediating device between people with IDD and others in their lives to remedy these problems as discussed by participants currently using data to manage their daily activities or the activities of those around them. However, visualizations must be able to adapt to the varying roles that data plays in each scenario. For Persona A, numeric data can be compared against concrete thresholds to understand aspects of treatment and provide agency in his own actions, privacy, and decisions. 
Persona B lacks access to the data-driven dialogues that happen about her between support staff and an ability to connect the data she tracks to the data used to characterize her well-being to her therapist. These observations raise key questions about transparency, privacy, awareness, collaboration, expression, and communication that offer open challenges for visualization design. 

 \subsection{\textbf{What Challenges Do You Face With Data?}} 
\subsubsection{Disability and data literacy limit accessibility.}  \ \ \label{sec:literacy}

\vspace{2pt}\noindent
Each individual with IDD has unique strengths and weaknesses, which may result in distinctly different experiences with data. Many people with IDD do not have the opportunity or motivation to improve their data literacy due to a lack of educational availability or their own perceived inabilities grounded in harmful stereotypes. Participants frequently reported that data can be difficult to understand and work with. For example, P2 used an analogy to describe how working with data can be highly specialized, 

\vspace{2pt}
\emph{"It's like when you go to a doctor and they're talking to you using med terms. You almost have to say, remember, you have to talk down to me. We're not at the same medical level you are at."} --- P2 (ID, middle literacy, high confidence)

\vspace{2pt} 
P11, a self-advocate who had extensive experience working with data as well as with people with IDD, mentioned that a lack of data literacy is a significant challenge for people in this community. He suggested that addressing literacy limitations should start with education and emphasized the importance of incorporating data education into everyday problem solving, 

\vspace{2pt}
\emph{"We're very good at teaching pass/fail situations, but not real life. Usually after you leave school, is there any other option for data? It's like you could live without data for a good majority of the time but not knowing it."} --- P11 (ID, high literacy, high confidence)

\vspace{2pt} 
While research frequently groups various disabilities together by necessity, different types of IDDs and levels of severity can also result in respective data processing difficulties. Participants with Autism reported being highly susceptible to cognitive overload and were more likely to become overstimulated with too much data. As mentioned by P13, 

\vspace{2pt}
\emph{"One of my issues is that, like with questions like this [anything else you want me to know about your experience with data], like I have so many things that come to mind all at once I call it, I call it like a bottleneck. It’s one of the issues that I have on a daily basis."} --- P13 (ASD, high literacy, middle confidence)

\vspace{2pt} 
Other IDDs make retaining information challenging. For example, P8 talked about cognitive challenges introduced by the Fetal Alcohol Syndrome,

\vspace{2pt}
\emph{"Part of my Fetal Alcohol Syndrome is that I'm missing some, sometimes someone will give me some information, it'll go in one ear and right out the other, or it will stay in short-term for a little bit. And I can't recall it because I have recalling problems. So then I get frustrated. So I think because of the amount of data that's going into my brain, the brain doesn't quite know how to handle it, a lot of it goes in and goes out and I forget about it."} --- P8 (Fetal Alcohol Syndrome \& BPD, high literacy, high confidence)

\vspace{2pt} 
P11 discussed the integration issues of trying to work with data from multiple sources, noting perspectives on both overload and retention. Specifically, they noted that tools might 
help people manage the heterogeneity of data they face every day, 

\vspace{2pt}
\emph{"My issue is integration of all of that data...It should be integrated and simple. You shouldn't be looking at five different apps just to figure out your day."} --- P11 (ID, high literacy, high confidence)

\vspace{2pt} 
Several participants also complained about data accuracy and expressed concerns about malicious data and misinformation. For example, P1 worried about editing technology, 

\vspace{2pt}
\emph{``I've seen a lot of people have a hard time understanding what they're viewing, because it can be traumatic to watch, some of these videos that happened, a lot of them are edited in a way that you don't know if it's even real because of the editing technology.''} --- P1 (ASD \& ADHD, high literacy, high confidence)

These observations highlighted several common challenges facing IDD community: (1) data is abstract and people with IDD typically have little to no educational training to make sense of it or discern its quality, (2) working with data is complicated and requires multiple cognitive skills such as retaining, recalling, and simultaneously integrating information which are areas in which people with IDD face significant challenges, and (3) current data education and analytics usually happen in a vacuum (e.g., in the classroom to solve an imaginary problem) rather than in real-life (e.g., to support personally relevant decision-making), which results in a lack of awareness of everyday data and low interest in engaging in data-driven activities. 

\subsubsection{Caregivers are important for data access.}  \ \ \label{sec:caregiver}

\vspace{2pt}\noindent
Severely impaired individuals may need full-time support or are otherwise unable to function on their own. Caregivers play an important role in their data access, sharing, and sensemaking. For these individuals, their sources of information usually are people rather than technology, meaning data pertaining to their daily lives is first interpreted by others. This lack of direct access significantly reduces their personal agency in data-driven reasoning and decision-making. As mentioned by P9's support staff, 

\vspace{2pt}
\emph{``Her go-to is not technology. Anything that she does is usually paper-pencil based. She gleans information from other sources other than technology. She and I get information together from the internet from technology, but on her own, that is not happening.''} --- on behalf of P9 (Williams Syndrome, low literacy, low confidence)

\vspace{2pt}
People with communication challenges are often unable to clearly express their needs and feelings, and are significantly limited in data reasoning processes. These situations become exacerbated when critical decisions are made by others on their behalf. \emph{``We just picked things that we thought he'd be interested in.''}, said the caregiver of P7 (Nonverbal ASD, low literacy, N/A confidence).

Less independent individuals are also vulnerable to exploitation by others. For example, a parent of a participant commented, 

\vspace{2pt}
\emph{``Our challenge is to teach her how to be more safe... She's doing all these different things to be able to interact in the world. But if a stranger walks up and says, give me your wallet, she'll give it to them. So it's just scary. Everyday is just trying to figure out the ups and downs of what to do.''} --- on behalf of P14 (Developmental Delay \& SPD, low literacy, low confidence)

\vspace{2pt}
While this example speaks to scenarios outside of data, the lack of data awareness exhibited by more severely impaired participants indicates vulnerabilities as well: people may not have access to or knowledge of the data they share or the consequences of that sharing. 

Further, it is also worth noting the data access needs and challenges for IDD advocates and caregivers to advocate on behalf of people with IDD. Throughout our interviews, data was considered as a form of authority and tool for storytelling for effective advocacy in healthcare, employment, education, and legal spaces. Participants appreciated the power of visualization to reinforce their messages and communicate positive narratives; however, they also noted a lack of visualization and data literacy, which is often not offered in most special education programs but needed to create data-driven graphics and narratives. As P7's caregiver stated, 

\vspace{2pt}
\emph{``I'm very quiet for myself, but when it comes to my children, I'm a very strong advocate. I feel like I didn't do justice for him because I wish I had known about inclusive education. I think that would have helped a lot with his development.''} 

These examples raised several ethical issues that people with IDD and their caregivers may encounter when interacting with data and offer critical design opportunities for visualization. For example, appropriate, co-designed data physicalization and other ``unplugged'' visualization techniques and activities (e.g., visualize personal routine, create a visual calendar, sort clothes by color/size/type, graphically portion control each meal and macronutrients, etc.) may create accessible gateways for people with IDD to improve data literacy, engage in data-driven activities, and raise data awareness in everyday decision-making by removing technological obstacles to working with data. In the cases where a person 
faces communicative challenges, collaborative visualization can encourage critical thinking and reflection between people with IDD and others in their lives, which may help create a shared conversational space between multiple parties and facilitate mutual understanding through data. 
\section{Discussion} 
Our interviews revealed that participants frequently used personal data for self-advocacy, self-expression, everyday function, and decision-making. However, due to a general lack of data literacy and stereotype-driven apathy, this data is usually invisible and even inaccessible to people with IDD, creating a cognitive and emotional barrier for them to actively participate in data-driven activities. In this section, we synthesize core findings and observations from the study and propose a near-term research agenda for cognitively accessible visualization to address these challenges and improve data accessibility for people with IDD. We discuss the needs for a collective effort to make data feel personal to motivate improved literacy and to design thoughtful data visualizations with people in mind.

\subsection{Make Data Feel Personal}
All of our participants were either self-tracking or being tracked by support staff. However, most participants and caregivers were unaware of the resulting data and often overlooked its role in their daily life. Data was usually associated with ``computer technology'' or ``hard drive,'' and generated a sense of personal insecurity and emotional distance. This lack of awareness and aversion to data can introduce personal challenges for this population. The tracked data reflected a range of personal situations for people with IDD but often was used to make decisions without their participation. In circumstances where they proactively collected data, they lacked 
sufficient means of generating meaning from the data and were sometimes unable to use it for expression, reflection, and advocacy. 

These challenges highlight a need for more inclusive visualization pedagogy for people with IDD. How to build such a pedagogy remains an important and open question. Past research demonstrated positive correlations between one's numeracy, need for cognition and visualization literacy ~\cite{Lee_Correlation_2019}, and discussed the need to situate literacy education in real life and develop an inclusive vocabulary ~\cite{visliteracy}. However, participants in our studies found the most significant challenge in data literacy development not to be in improving curricula but in helping overcome negative stereotypes and understanding how improved data literacy will benefit and empower them. Specifically, our interviews indicated two core challenges
for inclusive visualization: how can we surface the personal relevance of data for people with IDD,
and how can we design visualizations that are sensitive to individual differences and support personal meaning-making? 

Throughout the interviews, there was a correlation between one's confidence in acquiring data literacy and their level of 
personal motivation and past data experiences. An IDD self-advocate (P11) mentioned that improving one's own data literacy is a personal decision: many people do not see the need for working with data and hence are not actively seeking opportunities to increase their data literacy. While our findings illustrate a need for more accessible data science education, past literature in data science education suggests that without relevant frames and contexts, enhanced education alone is unlikely to be enough ~\cite{gebre2022conceptions}. Moreover, the harmful stereotype that data is something ``reserved for a computer technology kind of gal'' (P9) can further demotivate people to learn about data and claim agency. Here, we propose two considerations to make data feel personal, namely to promote data agency by removing visible technologies through physicalization and improve its relevance through personalization. 

\noindent\textbf{(1) Promote Data Agency through Physicalization:} 
Less independent participants often had many co-occurring health challenges and relied on tracking everyday routines (e.g., events, to-dos, health information, etc.) for functional autonomy. Our discussions with caregivers and people with IDD additionally highlighted accessibility and identity challenges associated with general technology use (see \S \ref{sec:apathy}). 
The easiest way for these individuals to access data was usually to use physical entities and everyday objects, such as a measuring cup, handwritten checklist, or nutrition label, that supports scaffolding and self-paced exploration. Low-fi data physicalization~\cite{jansen2015opportunities,huron2014constructive,bae2022making} and other 
``unplugged'' visualization techniques, such as Legos ~\cite{SurezGmez2020PhysicalVO}, kirigami ~\cite{kiriphys}, and knitting ~\cite{smit_2021}, may provide a more accessible means for people with IDD to develop a better data agency and engage with data. For example, creating visualizations with everyday materials through a constructionist approach~\cite{huron2014constructive} might help people with IDD create visualizations of daily activities or relevant tracking measures that could be used to communicate with caregivers or engage with data concepts in more familiar ways. When data is involved in these situations, visualization may serve as a boundary object ~\cite{mark2007boundary}---an object that bridges discrepancies in technical knowledge and power dynamics---crucial not only to help people with IDD better express themselves, but also to facilitate understanding and mutual agency. Moreover, tangible interactions, such as touch, smell and taste, may engage multiple senses and create an embodied experience for people with IDD to learn, reflect, and own their data.  \ \


\noindent\textbf{(2) Improve Data Relevance through Personalization:} More independent participants often deliberately collected data, particularly personal data (e.g., written notes, curated video clips, and everyday documentation) for self-reflection, expression, and advocacy. Such data becomes most meaningful 
when it is relevant to the individual and addresses a problem or achieves a goal of their personal interest, such as reflecting on and improving one's behavior (P1), shaping medical treatments for an individual or their pet (P6, P13, P15), and facilitating critical conversations around independence and competence between multiple stakeholders (P4, P7). However, working with these data and extracting meaning from it can be difficult. For many participants, all data is perceived as technical instead of personal, intimidating instead of exciting. Participants noted that ''the change should start with education,'' implying that earlier exposure to data and focus on literacy skills could help reverse systemic exclusionary patterns. Current data education curricula and analytics for children with IDD mostly focuses on idealized ``pass/fail situations'' and teaching to address hypothetical problems rather than discussing data in the context of real-life challenges. As a result, people often fail to see the connection between data and themselves, and can `` live without data for a good amount of time after leaving school'' (P11). Once 
people lose interest in data, ``it is hard to change their minds'' (P1). 

Research in CS education shows that teaching about topics like data visualization within the context of problems of personal or social relevance can increase engagement with and interest in those topics~\cite{kafai2020theory}. Integrating ideas from personal informatics into data science education may help provide contextually-meaningful scenarios for educators. 
Further, designing pedagogical approaches that attend to the complex and diverse skillsets of people with IDD may help communicate concepts relying on those skillsets more effectively. For example, providing different types of visual feedback, such as a regular chart or a living metaphor, may motivate 
people to engage with data in different ways due to differences in factors like working memory or numerical reasoning abilities~\cite{personaldata}. 
Future work in cognitively accessible visualization and inclusive data education should attend to individual differences and abilities, striving for better data relevance and encouraging personalized data discovery that suits individual preferences, needs, and skills. 


\subsection{Design Data Visualizations With People in Mind}
Visualization has a strong tradition of human-centered design approaches. However, our core practices were overwhelmingly developed with neurotypical and often highly numerate populations. 
Working with data can be cognitively demanding, and people with IDD are limited in many cognitive areas used in conventional visualization sensemaking processes, usually receive little relevant training at school, and, as a result, tend to have relatively low data literacy. 
The idea of ``data'' 
caused significant anxiety for many of our participants and led 
them to begin to withdraw from the conversation. 
 Even highly data-literate 
participants frequently reported experiencing ``cognitive overload'' (P13), had ``recall and remembering issues'' (P8), and found ``data integration'' (P11) challenging. 
This population also commonly experiences an invisible emotional challenge with data. Our interviews revealed that participants often link data to personal identity yet characterize data on a personal level.  The long-standing social stigma around intellectual disability may inevitably lead to self-stigma~\cite{werner_scior_2022}, creating detrimental effects on one's self-esteem and self-efficacy and resulting in frequent psychological distress when they experience frustration with data~\cite{self-stigma}. Seeing stereotypes within their data or struggling to work with a visualization perpetuated negative self-perceptions. 
Participants' stories suggested that these situations led them to further internalize negative views and attitudes, continue distancing themselves from data, and refuse opportunities to work with it. These challenges draw attention to two important research directions for cognitively accessible visualization: how can we design visualizations that make data less intimidating,
and how can visualization link data to identity in more positive ways?

\noindent\textbf{(1) Design More Approachable Visualizations:} 
Regardless of disability and severity, participants enjoyed stories communicated in a variety of formats on different platforms, such as TV shows, movies, video games, music, pictures, and photographs.
They often collected datasets from these sources to tell stories of their own to address the challenges introduced by their disabilities. For those with communication impairments and social anxiety, movies and video clips depicting real-world situations served as a rhetorical medium to express their feelings and thoughts, release the emotional distress introduced by their social impairments, and mediate discussions with psychiatrists or behavior coaches to figure out solutions to various social challenges. 
Storytelling was also consistently seen as an empowering tool to remove the stigma faced by people with IDD and used to present convincing evidence of independence. People used words, photos and pictures, or records of daily progress and accomplishments to demonstrate their abilities, needs and challenges. They used these self-assembled datasets to communicate their potential to overcome negative narratives and advocate for their own agency and participation in society. 
However, participants collected, curated, and presented this data manually. They lacked the tools to readily and efficiently construct personalized narratives for different situations. Participants with IDD 
frequently struggle with cognitive overload. Many approaches to working with multimedia or text data---the two most frequently described datasets curated by our participants---involve complex interfaces that may be overwhelming rather than supportive~\cite{borgo2011survey,kucher2015text}. 

Future research in data storytelling could inform novel approaches for accessible visualization creation and consumption by situating data within relevant contexts using approachable and engaging forms. Our interviews revealed that data became less overwhelming when it came as a narrative. Many participants mentioned how they would benefit from ``a step-by-step thing or like a video graphic'' (P13) to learn a new task. Wrapping data in a compelling format, such as data videos ~\cite{amini2015understanding}, data comics ~\cite{bach2017emerging}, or games ~\cite{rapp2018gamification},
may take advantage of their tendency to view data as content to support sensemaking (e.g., the story, tasks to be done, or decisions to be made) when framed with respect to the context in which the data is used. 
Photorealistic and multimedia data representations, such as pictures, photos, audio, and narrations, might augment more abstract marks to provide rich context for data-driven communication and to help people more readily associate data with other relevant stories in data to, for example, rapidly find relevant video segments to illustrate social challenges or relevant moments in complex video streams. Further, by drawing more deeply on narrative techniques applied in conventional media~~\cite{bradbury2020documentary}, visualization designers can develop new principles to create compelling data narratives that help make data more inclusive for people with IDD. 
 
\noindent\textbf{(2) Humanize Data with Creative Representations:} 
Participants frequently used self-assembled datasets for self-expression and as a proxy for delivering messages and experiences to and with other people. Our interviews suggest that having access to the rich context in which this data is collected is key to successful communication and accurate understanding. However, these data usually are stand-alone pieces and subject to individual interpretation. In addition, the data must be collected before it can be used to communicate, introducing a lag between when a situation arises and when data can support effective communication about that situation. 
Visualizations separate data from the context the data describes and too often pay insufficient attention to people that the data describes. Cognitively accessible visualization should humanize data, finding the connection between numbers in the graph and the people who are viewing it and being characterized by it. 
Designers can overcome these limitations by considering 
affective design elements ~\cite{lee2022affective}
or even the anthropomorphic characteristics of the visualization ~\cite{morais2020showing}.  
Future research should also investigate tools and techniques to empower people with IDD to create compelling data presentations to communicate aspects of their own life to others who may not share their experiences. 
These authoring tools should actively consider the link between data and identity to support people with IDD in constructing and/or maintaining a positive self-image rather than exacerbating negative stereotypes. 

Developing such tools may require expanded guidelines for personal visualization~\cite{huang2014personal}. 
To support inclusive visualization authoring, tools and expressive methods need to shift focus from the data to the people that the data represents, developing templates for communicating the relationships between multiple stakeholders, the challenges they are facing, and their similarities and differences in thinking and behaving in order to proactively identify critical phenomena (e.g., bias, prejudice, and gaps in mutual understanding). Visualization authoring tools for people with IDD should enable people to craft data stories starting from the story itself and drilling down into the data rather than starting from the data to build up a story from individual marks and statistics. Further, visualization should help link data to aspects of personal identity and help people with IDD foster a positive self-image by, for example, considering the semantics of different types of data in narrative templates or guidelines. Inclusive visualization authoring also opens opportunities for developing new techniques that aid authors in translating their personal goals, missions, and values into relatable visual elements such as shapes, colors, and icons. 
 
\subsection{Limitations \& Future Work}
Our work focuses on the general challenges in using data and visualization to inform a near-term research agenda for cognitively accessible visualization. Future work should more deeply explore different visualization types and guidelines to understand their efficacy. However, doing so requires a better understanding of the limited range of scenarios and representations that people with IDD feel comfortable discussing. Our study offers initial insights into the role that data plays for people with IDD, and provides a preliminary platform to enable such investigations.  

While this study covers a broad range of relationships between people and data, limitations of the current study provide several avenues for future work. First, we did not collect data on formal diagnoses. As noted in Section \S \ref{sec:method}, IDD tends to be difficult to diagnose and is accompanied by a range of 
co-occuring conditions. 
We followed existing best practices for working with this population ~\cite{chi}, including voluntary self-reporting and recruiting perspectives from a wide range of IDD. The lack of formal diagnoses prevents us from formally analyzing causal relationships between people with different IDD and may limit the generalizability of our results. However, the consistency of our primary observations across participants indicates that the experiences and themes from our work offer a promising preliminary characterization of key open challenges for cognitively accessible visualization.

We relied on participants' self-reported data literacy for our analysis and used a qualitative approach to understanding their experiences with data. This approach allowed participants to conceptualize data in different ways, shifting the subsequent discussions. These shifts may, as a result, lead to higher variance in our results than a more constrained or structured survey. However, existing approaches ~\cite{6875906, 7539634} to measuring data literacy rely on a level of detail and numeracy that our collaborators with IDD universally felt would be inappropriate for and even alienating to many people with IDD. This approach would prevent us from understanding the data needs of a large percentage of the community given existing attitudes towards data and the lack of data science education for people with IDD. Future work should further probe a broader population of people with IDD using more specific questions to understand the initial hypotheses and opportunities raised here. For example, future studies may explore literacy assessment approaches better tuned to the needs of people with IDD and how varying degrees of data literacy influence barriers to visualization use for people with IDD.

\section{Conclusion}
Conventional data visualization tools and guidelines often do not actively consider the specific needs and abilities of people with Intellectual and Developmental Disabilities (IDD), leaving them excluded from data-driven activities and vulnerable to ethical issues. In this paper, we conducted 15 semi-structured interviews with people with IDD and their caregivers to explore how people with IDD encounter data in everyday life and the role visualization may play in data accessibility. We categorized our findings into three distinct themes summarizing (1) how they conceptualize data; (2) when and where they encounter data; and (3) what challenges exist when they interact with data. Drawing on findings and observations, we synthesized core takeaways for visualization and 
identify key opportunities for accessible visualization. 
Our exploratory interviews can inform new concepts to be used to establish a research agenda for cognitively accessible visualization grounded in the authentic lived experiences of people with IDD.

\begin{acks}
We would like to thank Dr. Ben Shapiro and Dr. Shea Tanis for encouraging us to do this study. This work is funded by NSF \#2046725 and \#1933915,  as well as the Ray Hauser Award at University of Colorado Boulder. 
\end{acks}

\bibliographystyle{ACM-Reference-Format} 
\bibliography{idd-manuscript.bib}


\begin{thebibliography}{71}


\ifx \showCODEN    \undefined \def \showCODEN     #1{\unskip}     \fi
\ifx \showDOI      \undefined \def \showDOI       #1{#1}\fi
\ifx \showISBNx    \undefined \def \showISBNx     #1{\unskip}     \fi
\ifx \showISBNxiii \undefined \def \showISBNxiii  #1{\unskip}     \fi
\ifx \showISSN     \undefined \def \showISSN      #1{\unskip}     \fi
\ifx \showLCCN     \undefined \def \showLCCN      #1{\unskip}     \fi
\ifx \shownote     \undefined \def \shownote      #1{#1}          \fi
\ifx \showarticletitle \undefined \def \showarticletitle #1{#1}   \fi
\ifx \showURL      \undefined \def \showURL       {\relax}        \fi
\providecommand\bibfield[2]{#2}
\providecommand\bibinfo[2]{#2}
\providecommand\natexlab[1]{#1}
\providecommand\showeprint[2][]{arXiv:#2}

\bibitem[\protect\citeauthoryear{??}{sel}{2015}]%
        {self-stigma}
 \bibinfo{year}{2015}\natexlab{}.
\newblock \showarticletitle{Self-reported Stigma and Symptoms of Anxiety and
  Depression in People with Intellectual Disabilities: Findings from a Cross
  Sectional Study in England}.
\newblock \bibinfo{journal}{\emph{Journal of Affective Disorders}}
  \bibinfo{volume}{187} (\bibinfo{year}{2015}), \bibinfo{pages}{224--231}.
\newblock
\showISSN{0165-0327}
\urldef\tempurl%
\url{https://doi.org/10.1016/j.jad.2015.07.046}
\showDOI{\tempurl}


\bibitem[\protect\citeauthoryear{??}{bey}{2022}]%
        {beyondwords}
 \bibinfo{year}{2022}\natexlab{}.
\newblock \bibinfo{title}{Beyond Words}.
\newblock
\newblock
\urldef\tempurl%
\url{https://booksbeyondwords.co.uk}
\showURL{%
\tempurl}


\bibitem[\protect\citeauthoryear{??}{edu}{2022}]%
        {education}
 \bibinfo{year}{2022}\natexlab{}.
\newblock \bibinfo{title}{Issues, Disability Education}.
\newblock
\newblock
\urldef\tempurl%
\url{https://www.ncld.org/sigdispro/}
\showURL{%
\tempurl}


\bibitem[\protect\citeauthoryear{AAIDD}{AAIDD}{2022}]%
        {faqs}
\bibfield{author}{\bibinfo{person}{AAIDD}.} \bibinfo{year}{2022}\natexlab{}.
\newblock \bibinfo{title}{{F}{A}{Q}s on {I}ntellectual {D}isability ---
  aaidd.org}.
\newblock
  \bibinfo{howpublished}{\url{https://www.aaidd.org/intellectual-disability/faqs-on-intellectual-disability}}.
\newblock


\bibitem[\protect\citeauthoryear{Amini, Henry~Riche, Lee, Hurter, and
  Irani}{Amini et~al\mbox{.}}{2015}]%
        {amini2015understanding}
\bibfield{author}{\bibinfo{person}{Fereshteh Amini}, \bibinfo{person}{Nathalie
  Henry~Riche}, \bibinfo{person}{Bongshin Lee}, \bibinfo{person}{Christophe
  Hurter}, {and} \bibinfo{person}{Pourang Irani}.}
  \bibinfo{year}{2015}\natexlab{}.
\newblock \showarticletitle{Understanding Data Videos: Looking at Narrative
  Visualization through the Cinematography Lens}. In
  \bibinfo{booktitle}{\emph{Proceedings of the 33rd Annual ACM Conference on
  Human Factors in Computing Systems}}. \bibinfo{pages}{1459--1468}.
\newblock
\urldef\tempurl%
\url{https://doi.org/10.1145/2702123.2702431}
\showDOI{\tempurl}


\bibitem[\protect\citeauthoryear{Anderson}{Anderson}{1997}]%
        {anderson}
\bibfield{author}{\bibinfo{person}{M Anderson}.}
  \bibinfo{year}{1997}\natexlab{}.
\newblock \showarticletitle{Picture Activity Schedules and Engagement of Adults
  with Mental Retardation in A Group Home}.
\newblock \bibinfo{journal}{\emph{Research in Developmental Disabilities}}
  \bibinfo{volume}{18} (\bibinfo{date}{08} \bibinfo{year}{1997}),
  \bibinfo{pages}{231--250}.
\newblock
\urldef\tempurl%
\url{https://doi.org/10.1016/s0891-4222(97)00006-1}
\showDOI{\tempurl}


\bibitem[\protect\citeauthoryear{Association}{Association}{2013}]%
        {diagnostic}
\bibfield{author}{\bibinfo{person}{American~Psychiatric Association}.}
  \bibinfo{year}{2013}\natexlab{}.
\newblock \bibinfo{booktitle}{\emph{Diagnostic and Statistical Manual of Mental
  Disorders} (\bibinfo{edition}{5} ed.)}.
\newblock \bibinfo{publisher}{American Psychiatric Association}.
\newblock


\bibitem[\protect\citeauthoryear{Bach, Riche, Carpendale, and Pfister}{Bach
  et~al\mbox{.}}{2017}]%
        {bach2017emerging}
\bibfield{author}{\bibinfo{person}{Benjamin Bach},
  \bibinfo{person}{Nathalie~Henry Riche}, \bibinfo{person}{Sheelagh
  Carpendale}, {and} \bibinfo{person}{Hanspeter Pfister}.}
  \bibinfo{year}{2017}\natexlab{}.
\newblock \showarticletitle{The Emerging Genre of Data Comics}.
\newblock \bibinfo{journal}{\emph{IEEE Computer Graphics and Applications}}
  \bibinfo{volume}{37}, \bibinfo{number}{3} (\bibinfo{year}{2017}),
  \bibinfo{pages}{6--13}.
\newblock
\urldef\tempurl%
\url{https://doi.org/10.1109/MCG.2017.33}
\showDOI{\tempurl}


\bibitem[\protect\citeauthoryear{Bae, Zheng, West, Do, Huron, and Szafir}{Bae
  et~al\mbox{.}}{2022}]%
        {bae2022making}
\bibfield{author}{\bibinfo{person}{S~Sandra Bae}, \bibinfo{person}{Clement
  Zheng}, \bibinfo{person}{Mary~Etta West}, \bibinfo{person}{Ellen Yi-Luen Do},
  \bibinfo{person}{Samuel Huron}, {and} \bibinfo{person}{Danielle~Albers
  Szafir}.} \bibinfo{year}{2022}\natexlab{}.
\newblock \showarticletitle{Making Data Tangible: A Cross-disciplinary Design
  Space for Data Physicalization}. In \bibinfo{booktitle}{\emph{CHI Conference
  on Human Factors in Computing Systems}}. \bibinfo{pages}{1--18}.
\newblock
\urldef\tempurl%
\url{https://doi.org/10.1145/3491102.3501939}
\showDOI{\tempurl}


\bibitem[\protect\citeauthoryear{Baxter, Glendinning, and Clarke}{Baxter
  et~al\mbox{.}}{2008}]%
        {information}
\bibfield{author}{\bibinfo{person}{K Baxter}, \bibinfo{person}{C Glendinning},
  {and} \bibinfo{person}{S Clarke}.} \bibinfo{year}{2008}\natexlab{}.
\newblock \showarticletitle{Making Informed Choices in Social Care: the
  Importance of Accessible Information}.
\newblock \bibinfo{journal}{\emph{Health \& Social Care in the Community}}
  (\bibinfo{year}{2008}), \bibinfo{pages}{197--207}.
\newblock
\showISSN{0966-0410}
\urldef\tempurl%
\url{https://doi.org/10.1111/j.1365-2524.2007.00742.x}
\showDOI{\tempurl}


\bibitem[\protect\citeauthoryear{Boardman, Bernal, and Hollins}{Boardman
  et~al\mbox{.}}{2014}]%
        {boardman_bernal_hollins_2014}
\bibfield{author}{\bibinfo{person}{Liz Boardman}, \bibinfo{person}{Jane
  Bernal}, {and} \bibinfo{person}{Sheila Hollins}.}
  \bibinfo{year}{2014}\natexlab{}.
\newblock \showarticletitle{Communicating with People with Intellectual
  Disabilities: A Guide for General Psychiatrists}.
\newblock \bibinfo{journal}{\emph{Advances in Psychiatric Treatment}}
  \bibinfo{volume}{20} (\bibinfo{year}{2014}), \bibinfo{pages}{27–36}.
\newblock
\urldef\tempurl%
\url{https://doi.org/10.1192/apt.bp.110.008664}
\showDOI{\tempurl}


\bibitem[\protect\citeauthoryear{Boat, Wu, and (U.S.)}{Boat
  et~al\mbox{.}}{2015}]%
        {boat_2015_mental}
\bibfield{author}{\bibinfo{person}{Thomas~F Boat}, \bibinfo{person}{Joel~T Wu},
  {and} \bibinfo{person}{Institute Of~Medicine (U.S.)}.}
  \bibinfo{year}{2015}\natexlab{}.
\newblock \bibinfo{booktitle}{\emph{Mental Disorders and Disabilities Among
  Low-income Children}}.
\newblock \bibinfo{publisher}{National Academies Press}.
\newblock
\urldef\tempurl%
\url{https://doi.org/10.17226/21780}
\showDOI{\tempurl}


\bibitem[\protect\citeauthoryear{Borgo, Chen, Daubney, Grundy, Heidemann,
  H{\"o}ferlin, H{\"o}ferlin, J{\"a}nicke, Weiskopf, and Xie}{Borgo
  et~al\mbox{.}}{2011}]%
        {borgo2011survey}
\bibfield{author}{\bibinfo{person}{Rita Borgo}, \bibinfo{person}{Min Chen},
  \bibinfo{person}{Ben Daubney}, \bibinfo{person}{Edward Grundy},
  \bibinfo{person}{Gunther Heidemann}, \bibinfo{person}{Benjamin H{\"o}ferlin},
  \bibinfo{person}{Markus H{\"o}ferlin}, \bibinfo{person}{Heike J{\"a}nicke},
  \bibinfo{person}{Daniel Weiskopf}, {and} \bibinfo{person}{Xianghua Xie}.}
  \bibinfo{year}{2011}\natexlab{}.
\newblock \showarticletitle{A Survey on Video-based Graphics and Video
  Visualization.}
\newblock \bibinfo{journal}{\emph{Eurographics (State of the Art Reports)}}
  (\bibinfo{year}{2011}), \bibinfo{pages}{1--23}.
\newblock
\showISSN{1017-4656}
\urldef\tempurl%
\url{https://doi.org/10.2312/EG2011/stars/001-023}
\showDOI{\tempurl}


\bibitem[\protect\citeauthoryear{Borner and Maltese}{Borner and
  Maltese}{2015}]%
        {visliteracy}
\bibfield{author}{\bibinfo{person}{Katy Borner} {and} \bibinfo{person}{Adam
  Maltese}.} \bibinfo{year}{2015}\natexlab{}.
\newblock \showarticletitle{Investigating Aspects of Data Visualization
  Literacy Using 20 Information Visualizations and 273 Science Museum
  Visitors}.
\newblock \bibinfo{journal}{\emph{Information Visualization.}}
  (\bibinfo{date}{01} \bibinfo{year}{2015}), \bibinfo{pages}{1--16}.
\newblock


\bibitem[\protect\citeauthoryear{Bouras and Holt}{Bouras and Holt}{2007}]%
        {bouras_holt_2007}
\bibfield{author}{\bibinfo{person}{N. Bouras} {and} \bibinfo{person}{G. Holt}.}
  \bibinfo{year}{2007}\natexlab{}.
\newblock \bibinfo{booktitle}{\emph{Psychiatric and Behavioural Disorders in
  Intellectual and Developmental Disabilities} (\bibinfo{edition}{2} ed.)}.
\newblock \bibinfo{publisher}{Cambridge University Press}.
\newblock
\urldef\tempurl%
\url{https://doi.org/10.1017/CBO9780511543616}
\showDOI{\tempurl}


\bibitem[\protect\citeauthoryear{Boy, Rensink, Bertini, and Fekete}{Boy
  et~al\mbox{.}}{2014}]%
        {6875906}
\bibfield{author}{\bibinfo{person}{Jeremy Boy}, \bibinfo{person}{Ronald~A.
  Rensink}, \bibinfo{person}{Enrico Bertini}, {and}
  \bibinfo{person}{Jean-Daniel Fekete}.} \bibinfo{year}{2014}\natexlab{}.
\newblock \showarticletitle{A Principled Way of Assessing Visualization
  Literacy}.
\newblock \bibinfo{journal}{\emph{IEEE Transactions on Visualization and
  Computer Graphics}} \bibinfo{volume}{20}, \bibinfo{number}{12}
  (\bibinfo{year}{2014}), \bibinfo{pages}{1963--1972}.
\newblock
\urldef\tempurl%
\url{https://doi.org/10.1109/TVCG.2014.2346984}
\showDOI{\tempurl}


\bibitem[\protect\citeauthoryear{Boyle, Fox, Havercamp, and Zubler}{Boyle
  et~al\mbox{.}}{2020}]%
        {boyle_2020_the}
\bibfield{author}{\bibinfo{person}{Coleen~A. Boyle},
  \bibinfo{person}{Michael~H. Fox}, \bibinfo{person}{Susan~M. Havercamp}, {and}
  \bibinfo{person}{Jennifer Zubler}.} \bibinfo{year}{2020}\natexlab{}.
\newblock \showarticletitle{The Public Health Response to the COVID-19 Pandemic
  for People with Disabilities}.
\newblock \bibinfo{journal}{\emph{Disability and Health Journal}}
  \bibinfo{volume}{13} (\bibinfo{date}{07} \bibinfo{year}{2020}),
  \bibinfo{pages}{100943}.
\newblock
\urldef\tempurl%
\url{https://doi.org/10.1016/j.dhjo.2020.100943}
\showDOI{\tempurl}


\bibitem[\protect\citeauthoryear{Bradbury and Guadagno}{Bradbury and
  Guadagno}{2020}]%
        {bradbury2020documentary}
\bibfield{author}{\bibinfo{person}{Judd~D Bradbury} {and}
  \bibinfo{person}{Rosanna~E Guadagno}.} \bibinfo{year}{2020}\natexlab{}.
\newblock \showarticletitle{Documentary Narrative Visualization: Features and
  Modes of Documentary Film in Narrative Visualization}.
\newblock \bibinfo{journal}{\emph{Information Visualization}}
  \bibinfo{volume}{19}, \bibinfo{number}{4} (\bibinfo{year}{2020}),
  \bibinfo{pages}{339--352}.
\newblock
\urldef\tempurl%
\url{https://doi.org/10.1177/1473871620925071}
\showDOI{\tempurl}


\bibitem[\protect\citeauthoryear{Braddock, Hemp, Tanis, Wu, Haffer,
  Intellectual, and Disabilities}{Braddock et~al\mbox{.}}{2017}]%
        {braddock}
\bibfield{author}{\bibinfo{person}{David~L Braddock}, \bibinfo{person}{Richard
  Hemp}, \bibinfo{person}{Emily~Shea Tanis}, \bibinfo{person}{Jiang Wu},
  \bibinfo{person}{Laura Haffer}, \bibinfo{person}{American Association~On
  Intellectual}, {and} \bibinfo{person}{Developmental Disabilities}.}
  \bibinfo{year}{2017}\natexlab{}.
\newblock \bibinfo{booktitle}{\emph{The State of the States in Intellectual and
  Developmental Disabilities, 2017}}.
\newblock \bibinfo{publisher}{American Association On Intellectual And
  Developmental Disabilities}.
\newblock


\bibitem[\protect\citeauthoryear{Card, Mackinlay, and Shneiderman}{Card
  et~al\mbox{.}}{1999}]%
        {book}
\bibfield{author}{\bibinfo{person}{Stuart Card}, \bibinfo{person}{Jock
  Mackinlay}, {and} \bibinfo{person}{Ben Shneiderman}.}
  \bibinfo{year}{1999}\natexlab{}.
\newblock \bibinfo{booktitle}{\emph{Readings in Information Visualization:
  Using Vision To Think}}.
\newblock
\showISBNx{978-1-55860-533-6}


\bibitem[\protect\citeauthoryear{Carulla, Reed, Vaez-Azizi, Cooper, Leal,
  Bertelli, Adnams, Cooray, Deb, Dirani, et~al\mbox{.}}{Carulla
  et~al\mbox{.}}{2011}]%
        {carulla2011intellectual}
\bibfield{author}{\bibinfo{person}{Luis~Salvador Carulla},
  \bibinfo{person}{Geoffrey~M Reed}, \bibinfo{person}{Leila~M Vaez-Azizi},
  \bibinfo{person}{Sally-Ann Cooper}, \bibinfo{person}{Rafael~Martinez Leal},
  \bibinfo{person}{Marco Bertelli}, \bibinfo{person}{Colleen Adnams},
  \bibinfo{person}{Sherva Cooray}, \bibinfo{person}{Shoumitro Deb},
  \bibinfo{person}{Leyla~Akoury Dirani}, {et~al\mbox{.}}}
  \bibinfo{year}{2011}\natexlab{}.
\newblock \showarticletitle{Intellectual Developmental Disorders: Towards A New
  Name, Definition and Framework for “Mental Retardation/Intellectual
  Disability” in ICD-11}.
\newblock \bibinfo{journal}{\emph{World Psychiatry}} \bibinfo{volume}{10},
  \bibinfo{number}{3} (\bibinfo{year}{2011}), \bibinfo{pages}{175}.
\newblock
\urldef\tempurl%
\url{https://doi.org/10.1002/j.2051-5545.2011.tb00045.x}
\showDOI{\tempurl}


\bibitem[\protect\citeauthoryear{Chundury, Patnaik, Reyazuddin, Tang, Lazar,
  and Elmqvist}{Chundury et~al\mbox{.}}{2022}]%
        {Chundury}
\bibfield{author}{\bibinfo{person}{P. Chundury}, \bibinfo{person}{B. Patnaik},
  \bibinfo{person}{Y. Reyazuddin}, \bibinfo{person}{C. Tang},
  \bibinfo{person}{J. Lazar}, {and} \bibinfo{person}{N. Elmqvist}.}
  \bibinfo{year}{2022}\natexlab{}.
\newblock \showarticletitle{Towards Understanding Sensory Substitution for
  Accessible Visualization: An Interview Study}.
\newblock \bibinfo{journal}{\emph{IEEE Transactions on Visualization \&
  Computer Graphics}} \bibinfo{volume}{28}, \bibinfo{number}{01}
  (\bibinfo{date}{jan} \bibinfo{year}{2022}), \bibinfo{pages}{1084--1094}.
\newblock
\showISSN{1941-0506}
\urldef\tempurl%
\url{https://doi.org/10.1109/TVCG.2021.3114829}
\showDOI{\tempurl}


\bibitem[\protect\citeauthoryear{Conti, Ahamad, and Stasko}{Conti
  et~al\mbox{.}}{2005}]%
        {conti2005attacking}
\bibfield{author}{\bibinfo{person}{Gregory Conti}, \bibinfo{person}{Mustaque
  Ahamad}, {and} \bibinfo{person}{John Stasko}.}
  \bibinfo{year}{2005}\natexlab{}.
\newblock \showarticletitle{Attacking Information Visualization System
  Usability Overloading and Deceiving the Human}. In
  \bibinfo{booktitle}{\emph{Proceedings of the 2005 Symposium on Usable privacy
  and security}}. \bibinfo{pages}{89--100}.
\newblock
\urldef\tempurl%
\url{https://doi.org/10.1145/1073001.1073010}
\showDOI{\tempurl}


\bibitem[\protect\citeauthoryear{Daneshzand, Perin, and Carpendale}{Daneshzand
  et~al\mbox{.}}{2022}]%
        {kiriphys}
\bibfield{author}{\bibinfo{person}{Foroozan Daneshzand},
  \bibinfo{person}{Charles Perin}, {and} \bibinfo{person}{Sheelagh
  Carpendale}.} \bibinfo{year}{2022}\natexlab{}.
\newblock \showarticletitle{KiriPhys: Exploring New Data Physicalization
  Opportunities}.
\newblock \bibinfo{journal}{\emph{IEEE transactions on Visualization and
  Computer Graphics}}  \bibinfo{volume}{PP} (\bibinfo{date}{October}
  \bibinfo{year}{2022}).
\newblock
\showISSN{1077-2626}
\urldef\tempurl%
\url{https://doi.org/10.1109/tvcg.2022.3209365}
\showDOI{\tempurl}


\bibitem[\protect\citeauthoryear{Elavsky, Bennett, and Moritz}{Elavsky
  et~al\mbox{.}}{2022}]%
        {Chartability}
\bibfield{author}{\bibinfo{person}{Frank Elavsky}, \bibinfo{person}{Cynthia
  Bennett}, {and} \bibinfo{person}{Dominik Moritz}.}
  \bibinfo{year}{2022}\natexlab{}.
\newblock \showarticletitle{How Accessible is My Visualization? Evaluating
  Visualization Accessibility with Chartability}.
\newblock \bibinfo{journal}{\emph{Computer Graphics Forum}}
  \bibinfo{volume}{41}, \bibinfo{number}{3} (\bibinfo{year}{2022}),
  \bibinfo{pages}{57--70}.
\newblock
\urldef\tempurl%
\url{https://doi.org/10.1111/cgf.14522}
\showDOI{\tempurl}


\bibitem[\protect\citeauthoryear{Emerson, Hatton, Felce, and Murphy}{Emerson
  et~al\mbox{.}}{2001}]%
        {learning_disabilities}
\bibfield{author}{\bibinfo{person}{Eric Emerson}, \bibinfo{person}{Chris
  Hatton}, \bibinfo{person}{David Felce}, {and} \bibinfo{person}{G. Murphy}.}
  \bibinfo{year}{2001}\natexlab{}.
\newblock \bibinfo{booktitle}{\emph{Learning Disabilities : the Fundamental
  facts.}}
\newblock \bibinfo{publisher}{The Foundation for People with Learning
  Disabilities}.
\newblock


\bibitem[\protect\citeauthoryear{Fekete, van Wijk, Stasko, and North}{Fekete
  et~al\mbox{.}}{2008}]%
        {inbook}
\bibfield{author}{\bibinfo{person}{Jean-Daniel Fekete}, \bibinfo{person}{Jarke
  van Wijk}, \bibinfo{person}{John Stasko}, {and} \bibinfo{person}{Chris
  North}.} \bibinfo{year}{2008}\natexlab{}.
\newblock \bibinfo{booktitle}{\emph{The Value of Information Visualization}}.
  Vol.~\bibinfo{volume}{4950}.
\newblock \bibinfo{pages}{1--18}.
\newblock
\showISBNx{978-3-540-70955-8}
\urldef\tempurl%
\url{https://doi.org/10.1007/978-3-540-70956-5_1}
\showDOI{\tempurl}


\bibitem[\protect\citeauthoryear{Gebre}{Gebre}{2022}]%
        {gebre2022conceptions}
\bibfield{author}{\bibinfo{person}{Engida Gebre}.}
  \bibinfo{year}{2022}\natexlab{}.
\newblock \showarticletitle{Conceptions and Perspectives of Data Literacy in
  Secondary Education}.
\newblock \bibinfo{journal}{\emph{British Journal of Educational Technology}}
  (\bibinfo{year}{2022}).
\newblock
\urldef\tempurl%
\url{https://doi.org/10.1111/bjet.13246}
\showDOI{\tempurl}


\bibitem[\protect\citeauthoryear{Gibbons}{Gibbons}{2018}]%
        {gibbons_2018}
\bibfield{author}{\bibinfo{person}{Sarah Gibbons}.}
  \bibinfo{year}{2018}\natexlab{}.
\newblock \bibinfo{title}{Empathy Mapping: The First Step in Design Thinking}.
\newblock
\newblock
\urldef\tempurl%
\url{https://www.nngroup.com/articles/empathy-mapping/}
\showURL{%
\tempurl}


\bibitem[\protect\citeauthoryear{G\"{o}tzelmann}{G\"{o}tzelmann}{2018}]%
        {tactile}
\bibfield{author}{\bibinfo{person}{T. G\"{o}tzelmann}.}
  \bibinfo{year}{2018}\natexlab{}.
\newblock \showarticletitle{Visually Augmented Audio-Tactile Graphics for
  Visually Impaired People}.
\newblock \bibinfo{journal}{\emph{ACM Trans. Access. Comput.}}
  \bibinfo{volume}{11}, \bibinfo{number}{2}, Article \bibinfo{articleno}{8}
  (\bibinfo{date}{jun} \bibinfo{year}{2018}), \bibinfo{numpages}{31}~pages.
\newblock
\showISSN{1936-7228}
\urldef\tempurl%
\url{https://doi.org/10.1145/3186894}
\showDOI{\tempurl}


\bibitem[\protect\citeauthoryear{Harley}{Harley}{2015}]%
        {harley_2015}
\bibfield{author}{\bibinfo{person}{Aurora Harley}.}
  \bibinfo{year}{2015}\natexlab{}.
\newblock \bibinfo{title}{Personas Make Users Memorable for Product Team
  Members}.
\newblock
\newblock
\urldef\tempurl%
\url{https://www.nngroup.com/articles/persona/}
\showURL{%
\tempurl}


\bibitem[\protect\citeauthoryear{Hassiotis, Barron, and Hall}{Hassiotis
  et~al\mbox{.}}{2013}]%
        {Hassiotis2013-ge}
\bibfield{author}{\bibinfo{person}{Angela Hassiotis},
  \bibinfo{person}{Diana~Andrea Barron}, {and} \bibinfo{person}{Ian Hall}.}
  \bibinfo{year}{2013}\natexlab{}.
\newblock \bibinfo{booktitle}{\emph{Intellectual Disability Psychiatry: A
  Practical Handbook}}.
\newblock \bibinfo{publisher}{Wiley-Blackwell}, \bibinfo{address}{Hoboken, NJ}.
\newblock


\bibitem[\protect\citeauthoryear{Holloway, Marriott, Butler, and
  Reinders}{Holloway et~al\mbox{.}}{2019}]%
        {holloway20193d}
\bibfield{author}{\bibinfo{person}{Leona Holloway}, \bibinfo{person}{Kim
  Marriott}, \bibinfo{person}{Matthew Butler}, {and} \bibinfo{person}{Samuel
  Reinders}.} \bibinfo{year}{2019}\natexlab{}.
\newblock \showarticletitle{3D Printed Maps and Icons for Inclusion: Testing in
  the Wild by People Who Are Blind or Have Low Vision}. In
  \bibinfo{booktitle}{\emph{The 21st International ACM SIGACCESS Conference on
  Computers and Accessibility}}. \bibinfo{pages}{183--195}.
\newblock
\urldef\tempurl%
\url{https://doi.org/10.1145/3308561.3353790}
\showDOI{\tempurl}


\bibitem[\protect\citeauthoryear{Huang, Tory, Aseniero, Bartram, Bateman,
  Carpendale, Tang, and Woodbury}{Huang et~al\mbox{.}}{2014}]%
        {huang2014personal}
\bibfield{author}{\bibinfo{person}{Dandan Huang}, \bibinfo{person}{Melanie
  Tory}, \bibinfo{person}{Bon~Adriel Aseniero}, \bibinfo{person}{Lyn Bartram},
  \bibinfo{person}{Scott Bateman}, \bibinfo{person}{Sheelagh Carpendale},
  \bibinfo{person}{Anthony Tang}, {and} \bibinfo{person}{Robert Woodbury}.}
  \bibinfo{year}{2014}\natexlab{}.
\newblock \showarticletitle{Personal Visualization and Personal Visual
  Analytics}.
\newblock \bibinfo{journal}{\emph{IEEE Transactions on Visualization and
  Computer Graphics}} \bibinfo{volume}{21}, \bibinfo{number}{3}
  (\bibinfo{year}{2014}), \bibinfo{pages}{420--433}.
\newblock
\urldef\tempurl%
\url{https://doi.org/10.1109/TVCG.2014.2359887}
\showDOI{\tempurl}


\bibitem[\protect\citeauthoryear{Huron, Carpendale, Thudt, Tang, and
  Mauerer}{Huron et~al\mbox{.}}{2014}]%
        {huron2014constructive}
\bibfield{author}{\bibinfo{person}{Samuel Huron}, \bibinfo{person}{Sheelagh
  Carpendale}, \bibinfo{person}{Alice Thudt}, \bibinfo{person}{Anthony Tang},
  {and} \bibinfo{person}{Michael Mauerer}.} \bibinfo{year}{2014}\natexlab{}.
\newblock \showarticletitle{Constructive visualization}. In
  \bibinfo{booktitle}{\emph{Proceedings of the 2014 conference on Designing
  interactive systems}}. \bibinfo{pages}{433--442}.
\newblock
\urldef\tempurl%
\url{https://doi.org/10.1145/2598510.2598566}
\showDOI{\tempurl}


\bibitem[\protect\citeauthoryear{Issues}{Issues}{2022}]%
        {easyread}
\bibfield{author}{\bibinfo{person}{Office for~Disability Issues}.}
  \bibinfo{year}{2022}\natexlab{}.
\newblock \bibinfo{title}{A Guide to Making Easy Read information}.
\newblock
\newblock
\urldef\tempurl%
\url{https://www.odi.govt.nz/guidance-and-resources/a-guide-to-making-easy-read-information/}
\showURL{%
\tempurl}


\bibitem[\protect\citeauthoryear{Jansen, Dragicevic, Isenberg, Alexander,
  Karnik, Kildal, Subramanian, and Hornb{\ae}k}{Jansen et~al\mbox{.}}{2015}]%
        {jansen2015opportunities}
\bibfield{author}{\bibinfo{person}{Yvonne Jansen}, \bibinfo{person}{Pierre
  Dragicevic}, \bibinfo{person}{Petra Isenberg}, \bibinfo{person}{Jason
  Alexander}, \bibinfo{person}{Abhijit Karnik}, \bibinfo{person}{Johan Kildal},
  \bibinfo{person}{Sriram Subramanian}, {and} \bibinfo{person}{Kasper
  Hornb{\ae}k}.} \bibinfo{year}{2015}\natexlab{}.
\newblock \showarticletitle{Opportunities and Challenges for Data
  Physicalization}. In \bibinfo{booktitle}{\emph{Proceedings of the 33rd Annual
  ACM Conference on Human Factors in Computing Systems}}.
  \bibinfo{pages}{3227--3236}.
\newblock
\urldef\tempurl%
\url{https://doi.org/10.1145/2702123.2702180}
\showDOI{\tempurl}


\bibitem[\protect\citeauthoryear{Kafai, Proctor, and Lui}{Kafai
  et~al\mbox{.}}{2020}]%
        {kafai2020theory}
\bibfield{author}{\bibinfo{person}{Yasmin Kafai}, \bibinfo{person}{Chris
  Proctor}, {and} \bibinfo{person}{Debora Lui}.}
  \bibinfo{year}{2020}\natexlab{}.
\newblock \showarticletitle{From Theory Vias to Theory Dialogue: Embracing
  Cognitive, Situated, and Critical Framings of Computational Thinking in K-12
  CS Education}.
\newblock \bibinfo{journal}{\emph{ACM Inroads}} \bibinfo{volume}{11},
  \bibinfo{number}{1} (\bibinfo{year}{2020}), \bibinfo{pages}{44--53}.
\newblock
\urldef\tempurl%
\url{https://doi.org/10.1145/3291279.3339400}
\showDOI{\tempurl}


\bibitem[\protect\citeauthoryear{Kim, Joyner, Riegelhuth, and Kim}{Kim
  et~al\mbox{.}}{2021}]%
        {Kim2021AccessibleVD}
\bibfield{author}{\bibinfo{person}{Nam~Wook Kim},
  \bibinfo{person}{Shakila~Cherise Joyner}, \bibinfo{person}{Amalia
  Riegelhuth}, {and} \bibinfo{person}{Yea-Seul Kim}.}
  \bibinfo{year}{2021}\natexlab{}.
\newblock \showarticletitle{Accessible Visualization: Design Space,
  Opportunities, and Challenges}.
\newblock \bibinfo{journal}{\emph{Computer Graphics Forum}}
  \bibinfo{volume}{40} (\bibinfo{year}{2021}).
\newblock
\urldef\tempurl%
\url{https://doi.org/10.1111/cgf.14298}
\showDOI{\tempurl}


\bibitem[\protect\citeauthoryear{Kucher and Kerren}{Kucher and Kerren}{2015}]%
        {kucher2015text}
\bibfield{author}{\bibinfo{person}{Kostiantyn Kucher} {and}
  \bibinfo{person}{Andreas Kerren}.} \bibinfo{year}{2015}\natexlab{}.
\newblock \showarticletitle{Text Visualization Techniques: Taxonomy, Visual
  Survey, and Community Insights}. In \bibinfo{booktitle}{\emph{2015 IEEE
  Pacific Visualization Symposium (PacificVis)}}. IEEE,
  \bibinfo{pages}{117--121}.
\newblock
\urldef\tempurl%
\url{https://doi.org/10.1109/PACIFICVIS.2015.7156366}
\showDOI{\tempurl}


\bibitem[\protect\citeauthoryear{Lee, Choe, Isenberg, Marriott, and Stasko}{Lee
  et~al\mbox{.}}{2020}]%
        {broader}
\bibfield{author}{\bibinfo{person}{Bongshin Lee}, \bibinfo{person}{Eun~Kyoung
  Choe}, \bibinfo{person}{Petra Isenberg}, \bibinfo{person}{Kim Marriott},
  {and} \bibinfo{person}{John Stasko}.} \bibinfo{year}{2020}\natexlab{}.
\newblock \showarticletitle{Reaching Broader Audiences With Data
  Visualization}.
\newblock \bibinfo{journal}{\emph{IEEE Computer Graphics and Applications}}
  \bibinfo{volume}{40}, \bibinfo{number}{2} (\bibinfo{year}{2020}),
  \bibinfo{pages}{82--90}.
\newblock
\urldef\tempurl%
\url{https://doi.org/10.1109/MCG.2020.2968244}
\showDOI{\tempurl}


\bibitem[\protect\citeauthoryear{Lee, Isaacs, Szafir, Marai, Turkay, Tory,
  Carpendale, and Endert}{Lee et~al\mbox{.}}{2019a}]%
        {diversity}
\bibfield{author}{\bibinfo{person}{Bongshin Lee}, \bibinfo{person}{Kate
  Isaacs}, \bibinfo{person}{Danielle~Albers Szafir}, \bibinfo{person}{G.~E.
  Marai}, \bibinfo{person}{Cagatay Turkay}, \bibinfo{person}{Melanie Tory},
  \bibinfo{person}{Sheelagh Carpendale}, {and} \bibinfo{person}{Alex Endert}.}
  \bibinfo{year}{2019}\natexlab{a}.
\newblock \showarticletitle{Broadening Intellectual Diversity in Visualization
  Research Papers}.
\newblock \bibinfo{journal}{\emph{IEEE Computer Graphics and Applications}}
  \bibinfo{volume}{39}, \bibinfo{number}{4} (\bibinfo{year}{2019}),
  \bibinfo{pages}{78--85}.
\newblock
\urldef\tempurl%
\url{https://doi.org/10.1109/MCG.2019.2914844}
\showDOI{\tempurl}


\bibitem[\protect\citeauthoryear{Lee, Kim, and Kwon}{Lee et~al\mbox{.}}{2017}]%
        {7539634}
\bibfield{author}{\bibinfo{person}{Sukwon Lee}, \bibinfo{person}{Sung-Hee Kim},
  {and} \bibinfo{person}{Bum~Chul Kwon}.} \bibinfo{year}{2017}\natexlab{}.
\newblock \showarticletitle{VLAT: Development of a Visualization Literacy
  Assessment Test}.
\newblock \bibinfo{journal}{\emph{IEEE Transactions on Visualization and
  Computer Graphics}} \bibinfo{volume}{23}, \bibinfo{number}{1}
  (\bibinfo{date}{Jan} \bibinfo{year}{2017}), \bibinfo{pages}{551--560}.
\newblock
\showISSN{1941-0506}
\urldef\tempurl%
\url{https://doi.org/10.1109/TVCG.2016.2598920}
\showDOI{\tempurl}


\bibitem[\protect\citeauthoryear{Lee, Kwon, Yang, Lee, and Kim}{Lee
  et~al\mbox{.}}{2019b}]%
        {Lee_Correlation_2019}
\bibfield{author}{\bibinfo{person}{Sukwon Lee}, \bibinfo{person}{Bum~Chul
  Kwon}, \bibinfo{person}{Jiming Yang}, \bibinfo{person}{Byung~Cheol Lee},
  {and} \bibinfo{person}{Sung-Hee Kim}.} \bibinfo{year}{2019}\natexlab{b}.
\newblock \showarticletitle{The Correlation between Users’ Cognitive
  Characteristics and Visualization Literacy}.
\newblock \bibinfo{journal}{\emph{Applied Sciences}} \bibinfo{volume}{9},
  \bibinfo{number}{3} (\bibinfo{year}{2019}).
\newblock
\showISSN{2076-3417}
\urldef\tempurl%
\url{https://doi.org/10.3390/app9030488}
\showDOI{\tempurl}


\bibitem[\protect\citeauthoryear{Lee-Robbins and Adar}{Lee-Robbins and
  Adar}{2023}]%
        {lee2022affective}
\bibfield{author}{\bibinfo{person}{Elsie Lee-Robbins} {and}
  \bibinfo{person}{Eytan Adar}.} \bibinfo{year}{2023}\natexlab{}.
\newblock \showarticletitle{Affective Learning Objectives for Communicative
  Visualizations}.
\newblock \bibinfo{journal}{\emph{IEEE Transactions on Visualization and
  Computer Graphics}} (\bibinfo{year}{2023}).
\newblock
\urldef\tempurl%
\url{https://doi.org/10.1109/TVCG.2022.3209500}
\showDOI{\tempurl}


\bibitem[\protect\citeauthoryear{Lundgard, Lee, and Satyanarayan}{Lundgard
  et~al\mbox{.}}{2019}]%
        {lundgard2019sociotechnical}
\bibfield{author}{\bibinfo{person}{Alan Lundgard}, \bibinfo{person}{Crystal
  Lee}, {and} \bibinfo{person}{Arvind Satyanarayan}.}
  \bibinfo{year}{2019}\natexlab{}.
\newblock \showarticletitle{Sociotechnical considerations for accessible
  visualization design}. In \bibinfo{booktitle}{\emph{2019 IEEE Visualization
  Conference (VIS)}}. IEEE, \bibinfo{pages}{16--20}.
\newblock
\urldef\tempurl%
\url{https://doi.org/10.1109/VISUAL.2019.8933762}
\showDOI{\tempurl}


\bibitem[\protect\citeauthoryear{Lundgard and Satyanarayan}{Lundgard and
  Satyanarayan}{2022}]%
        {text-model}
\bibfield{author}{\bibinfo{person}{Alan Lundgard} {and} \bibinfo{person}{Arvind
  Satyanarayan}.} \bibinfo{year}{2022}\natexlab{}.
\newblock \showarticletitle{Accessible Visualization via Natural Language
  Descriptions: A Four-Level Model of Semantic Content}.
\newblock \bibinfo{journal}{\emph{IEEE Transactions on Visualization \&
  Computer Graphics (Proc. IEEE VIS)}} (\bibinfo{year}{2022}).
\newblock
\urldef\tempurl%
\url{https://doi.org/10.1109/TVCG.2021.3114770}
\showDOI{\tempurl}


\bibitem[\protect\citeauthoryear{Mark, Lyytinen, and Bergman}{Mark
  et~al\mbox{.}}{2007}]%
        {mark2007boundary}
\bibfield{author}{\bibinfo{person}{Gloria Mark}, \bibinfo{person}{Kalle
  Lyytinen}, {and} \bibinfo{person}{Mark Bergman}.}
  \bibinfo{year}{2007}\natexlab{}.
\newblock \showarticletitle{Boundary Objects in Design: An Ecological View of
  Design Artifacts}.
\newblock \bibinfo{journal}{\emph{Journal of the Association for Information
  Systems}} \bibinfo{volume}{8}, \bibinfo{number}{11} (\bibinfo{year}{2007}),
  \bibinfo{pages}{34}.
\newblock
\urldef\tempurl%
\url{https://doi.org/10.17705/1jais.00144}
\showDOI{\tempurl}


\bibitem[\protect\citeauthoryear{Marriott, Lee, Butler, Cutrell, Ellis, Goncu,
  Hearst, McCoy, and Szafir}{Marriott et~al\mbox{.}}{2021}]%
        {disability}
\bibfield{author}{\bibinfo{person}{Kim Marriott}, \bibinfo{person}{Bongshin
  Lee}, \bibinfo{person}{Matthew Butler}, \bibinfo{person}{Ed Cutrell},
  \bibinfo{person}{Kirsten Ellis}, \bibinfo{person}{Cagatay Goncu},
  \bibinfo{person}{Marti Hearst}, \bibinfo{person}{Kathleen McCoy}, {and}
  \bibinfo{person}{Danielle~Albers Szafir}.} \bibinfo{year}{2021}\natexlab{}.
\newblock \showarticletitle{Inclusive Data Visualization for People with
  Disabilities: A Call to Action}.
\newblock \bibinfo{journal}{\emph{Interactions}} \bibinfo{volume}{28},
  \bibinfo{number}{3} (\bibinfo{date}{apr} \bibinfo{year}{2021}),
  \bibinfo{pages}{47–51}.
\newblock
\showISSN{1072-5520}
\urldef\tempurl%
\url{https://doi.org/10.1145/3457875}
\showDOI{\tempurl}


\bibitem[\protect\citeauthoryear{Marrus and Hall}{Marrus and Hall}{2017}]%
        {marrus_2017_intellectual}
\bibfield{author}{\bibinfo{person}{Natasha Marrus} {and} \bibinfo{person}{Lacey
  Hall}.} \bibinfo{year}{2017}\natexlab{}.
\newblock \showarticletitle{Intellectual Disability and Language Disorder}.
\newblock \bibinfo{journal}{\emph{Child and Adolescent Psychiatric Clinics of
  North America}}  \bibinfo{volume}{26} (\bibinfo{date}{07}
  \bibinfo{year}{2017}), \bibinfo{pages}{539--554}.
\newblock
\urldef\tempurl%
\url{https://doi.org/10.1016/j.chc.2017.03.001}
\showDOI{\tempurl}


\bibitem[\protect\citeauthoryear{Montague}{Montague}{1997}]%
        {math}
\bibfield{author}{\bibinfo{person}{Marjorie Montague}.}
  \bibinfo{year}{1997}\natexlab{}.
\newblock \showarticletitle{Cognitive Strategy Instruction in Mathematics for
  Students with Learning Disabilities}.
\newblock \bibinfo{journal}{\emph{Journal of Learning Disabilities}}
  \bibinfo{volume}{30}, \bibinfo{number}{2} (\bibinfo{year}{1997}),
  \bibinfo{pages}{164--177}.
\newblock
\urldef\tempurl%
\url{https://doi.org/10.1177/002221949703000204}
\showDOI{\tempurl}


\bibitem[\protect\citeauthoryear{Morais, Jansen, Andrade, and
  Dragicevic}{Morais et~al\mbox{.}}{2022}]%
        {morais2020showing}
\bibfield{author}{\bibinfo{person}{Luiz Morais}, \bibinfo{person}{Yvonne
  Jansen}, \bibinfo{person}{Nazareno Andrade}, {and} \bibinfo{person}{Pierre
  Dragicevic}.} \bibinfo{year}{2022}\natexlab{}.
\newblock \showarticletitle{Showing Data about People: A Design Space of
  Anthropographics}.
\newblock \bibinfo{journal}{\emph{IEEE Transactions on Visualization and
  Computer Graphics}} (\bibinfo{year}{2022}).
\newblock
\urldef\tempurl%
\url{https://doi.org/10.1109/TVCG.2020.3023013}
\showDOI{\tempurl}


\bibitem[\protect\citeauthoryear{O'Kane and Goldbart}{O'Kane and
  Goldbart}{2016}]%
        {OKane2016-dl}
\bibfield{author}{\bibinfo{person}{Judith~Coupe O'Kane} {and}
  \bibinfo{person}{Juliet Goldbart}.} \bibinfo{year}{2016}\natexlab{}.
\newblock \bibinfo{booktitle}{\emph{Communication Before Speech: Development
  and Assessment}}.
\newblock \bibinfo{publisher}{CRC Press}, \bibinfo{address}{London, England}.
\newblock
\urldef\tempurl%
\url{https://doi.org/10.4324/9780203462157}
\showDOI{\tempurl}


\bibitem[\protect\citeauthoryear{Pernice}{Pernice}{2018}]%
        {pernice_2018}
\bibfield{author}{\bibinfo{person}{Kara Pernice}.}
  \bibinfo{year}{2018}\natexlab{}.
\newblock \bibinfo{title}{Affinity Diagramming: Collaboratively Sort UX
  Findings \& Design Ideas}.
\newblock
\newblock
\urldef\tempurl%
\url{https://www.nngroup.com/articles/affinity-diagram/}
\showURL{%
\tempurl}


\bibitem[\protect\citeauthoryear{Rapp}{Rapp}{2018}]%
        {rapp2018gamification}
\bibfield{author}{\bibinfo{person}{Amon Rapp}.}
  \bibinfo{year}{2018}\natexlab{}.
\newblock \showarticletitle{Gamification for Self-tracking: From World of
  Warcraft to the Design of Personal Informatics Systems}. In
  \bibinfo{booktitle}{\emph{Proceedings of the 2018 CHI Conference on Human
  Factors in Computing Systems}}. \bibinfo{pages}{1--15}.
\newblock
\urldef\tempurl%
\url{https://doi.org/10.1145/3173574.3173654}
\showDOI{\tempurl}


\bibitem[\protect\citeauthoryear{Sarju}{Sarju}{2021}]%
        {sarju_2021_nothing}
\bibfield{author}{\bibinfo{person}{Julia~P. Sarju}.}
  \bibinfo{year}{2021}\natexlab{}.
\newblock \showarticletitle{Nothing About Us Without Us – Towards Genuine
  Inclusion of Disabled Scientists and Science Students Post Pandemic}.
\newblock \bibinfo{journal}{\emph{Chemistry – A European Journal}}
  (\bibinfo{date}{06} \bibinfo{year}{2021}).
\newblock
\urldef\tempurl%
\url{https://doi.org/10.1002/chem.202100268}
\showDOI{\tempurl}


\bibitem[\protect\citeauthoryear{Schneider, Schauer, Stachl, and
  Butz}{Schneider et~al\mbox{.}}{2017}]%
        {personaldata}
\bibfield{author}{\bibinfo{person}{Hanna Schneider}, \bibinfo{person}{Katrin
  Schauer}, \bibinfo{person}{Clemens Stachl}, {and} \bibinfo{person}{Andreas
  Butz}.} \bibinfo{year}{2017}\natexlab{}.
\newblock \showarticletitle{Your Data, Your Vis: Personalizing Personal Data
  Visualizations}. \bibinfo{pages}{374--392}.
\newblock
\showISBNx{978-3-319-67686-9}
\urldef\tempurl%
\url{https://doi.org/10.1007/978-3-319-67687-6_25}
\showDOI{\tempurl}


\bibitem[\protect\citeauthoryear{Shogren, Kennedy, Dowsett, and Little}{Shogren
  et~al\mbox{.}}{2014}]%
        {shogren_2014_autonomy}
\bibfield{author}{\bibinfo{person}{Karrie~A. Shogren}, \bibinfo{person}{William
  Kennedy}, \bibinfo{person}{Chantelle Dowsett}, {and} \bibinfo{person}{Todd~D.
  Little}.} \bibinfo{year}{2014}\natexlab{}.
\newblock \showarticletitle{Autonomy, Psychological Empowerment, and
  Self-Realization: Exploring Data on Self-Determination from NLTS2}.
\newblock \bibinfo{journal}{\emph{Exceptional Children}}  \bibinfo{volume}{80}
  (\bibinfo{date}{01} \bibinfo{year}{2014}), \bibinfo{pages}{221--235}.
\newblock
\urldef\tempurl%
\url{https://doi.org/10.1177/001440291408000206}
\showDOI{\tempurl}


\bibitem[\protect\citeauthoryear{Smit}{Smit}{2021}]%
        {smit_2021}
\bibfield{author}{\bibinfo{person}{Noeska Smit}.}
  \bibinfo{year}{2021}\natexlab{}.
\newblock \bibinfo{title}{Data Knitualization: An Exploration of Knitting as a
  Visualization Medium}.
\newblock
\newblock
\urldef\tempurl%
\url{https://doi.org/10.31219/osf.io/xahj9}
\showDOI{\tempurl}


\bibitem[\protect\citeauthoryear{South and Borkin}{South and Borkin}{2023}]%
        {south}
\bibfield{author}{\bibinfo{person}{Laura South} {and}
  \bibinfo{person}{Michelle~A. Borkin}.} \bibinfo{year}{2023}\natexlab{}.
\newblock \showarticletitle{Photosensitive Accessibility for Interactive Data
  Visualizations}.
\newblock \bibinfo{journal}{\emph{IEEE Transactions on Visualization and
  Computer Graphics}} \bibinfo{volume}{29}, \bibinfo{number}{01}
  (\bibinfo{year}{2023}).
\newblock
\showISSN{1941-0506}
\urldef\tempurl%
\url{https://doi.org/10.1109/TVCG.2022.3209359}
\showDOI{\tempurl}


\bibitem[\protect\citeauthoryear{Sturmey and Didden}{Sturmey and
  Didden}{2014}]%
        {sturmey_2014_evidencebased}
\bibfield{author}{\bibinfo{person}{Peter Sturmey} {and} \bibinfo{person}{Robert
  Didden}.} \bibinfo{year}{2014}\natexlab{}.
\newblock \bibinfo{booktitle}{\emph{Evidence-based Practice and Intellectual
  Disabilities}}.
\newblock \bibinfo{publisher}{Wiley-Blackwell}.
\newblock


\bibitem[\protect\citeauthoryear{Su{\'a}rez-G{\'o}mez and
  Per{\'e}z-Holgu{\'i}n}{Su{\'a}rez-G{\'o}mez and
  Per{\'e}z-Holgu{\'i}n}{2020}]%
        {SurezGmez2020PhysicalVO}
\bibfield{author}{\bibinfo{person}{Andr{\'e}s-David Su{\'a}rez-G{\'o}mez} {and}
  \bibinfo{person}{Wilson~Javier Per{\'e}z-Holgu{\'i}n}.}
  \bibinfo{year}{2020}\natexlab{}.
\newblock \showarticletitle{Physical Visualization of Math Concepts Using LEGO
  Mindstorms}.
\newblock \bibinfo{journal}{\emph{Journal of Technology and Science Education}}
   \bibinfo{volume}{10} (\bibinfo{year}{2020}), \bibinfo{pages}{72--86}.
\newblock
\urldef\tempurl%
\url{https://doi.org/10.3926/jotse.788}
\showDOI{\tempurl}


\bibitem[\protect\citeauthoryear{Szymanski}{Szymanski}{2002}]%
        {psychiatric}
\bibfield{author}{\bibinfo{person}{L.~S. Szymanski}.}
  \bibinfo{year}{2002}\natexlab{}.
\newblock \showarticletitle{Diagnostic Criteria for Psychiatric Disorders for
  Use with Adults with Learning Disabilities/Mental Retardation}.
\newblock \bibinfo{journal}{\emph{Journal of Intellectual Disability Research}}
  \bibinfo{volume}{46}, \bibinfo{number}{6} (\bibinfo{year}{2002}),
  \bibinfo{pages}{525--527}.
\newblock
\urldef\tempurl%
\url{https://doi.org/10.1046/j.1365-2788.47.s1.25.x}
\showDOI{\tempurl}


\bibitem[\protect\citeauthoryear{Terras, Jarrett, and McGregor}{Terras
  et~al\mbox{.}}{2021}]%
        {accessibleinfo}
\bibfield{author}{\bibinfo{person}{Melody~M. Terras}, \bibinfo{person}{Dominic
  Jarrett}, {and} \bibinfo{person}{Sharon~A. McGregor}.}
  \bibinfo{year}{2021}\natexlab{}.
\newblock \showarticletitle{The Importance of Accessible Information in
  Promoting the Inclusion of People with an Intellectual Disability}.
\newblock \bibinfo{journal}{\emph{Disabilities}} \bibinfo{volume}{1},
  \bibinfo{number}{3} (\bibinfo{year}{2021}), \bibinfo{pages}{132--150}.
\newblock
\showISSN{2673-7272}
\urldef\tempurl%
\url{https://doi.org/10.3390/disabilities1030011}
\showDOI{\tempurl}


\bibitem[\protect\citeauthoryear{Tuffrey-Wijne, Bernal, Jones, Butler, and
  Hollins}{Tuffrey-Wijne et~al\mbox{.}}{2006}]%
        {tuffreywijne_2006_people}
\bibfield{author}{\bibinfo{person}{Irene Tuffrey-Wijne}, \bibinfo{person}{Jane
  Bernal}, \bibinfo{person}{Amelia Jones}, \bibinfo{person}{Gary Butler}, {and}
  \bibinfo{person}{Sheila Hollins}.} \bibinfo{year}{2006}\natexlab{}.
\newblock \showarticletitle{People with Intellectual Disabilities and Their
  Need for Cancer Information}.
\newblock \bibinfo{journal}{\emph{European Journal of Oncology Nursing}}
  \bibinfo{volume}{10} (\bibinfo{date}{04} \bibinfo{year}{2006}),
  \bibinfo{pages}{106--116}.
\newblock
\urldef\tempurl%
\url{https://doi.org/10.1016/j.ejon.2005.05.005}
\showDOI{\tempurl}


\bibitem[\protect\citeauthoryear{Van~Garderen}{Van~Garderen}{2006}]%
        {article}
\bibfield{author}{\bibinfo{person}{Delinda Van~Garderen}.}
  \bibinfo{year}{2006}\natexlab{}.
\newblock \showarticletitle{Spatial Visualization, Visual Imagery, and
  Mathematical Problem Solving of Students With Varying Abilities}.
\newblock \bibinfo{journal}{\emph{Journal of learning disabilities}}
  \bibinfo{volume}{39} (\bibinfo{date}{12} \bibinfo{year}{2006}),
  \bibinfo{pages}{496--506}.
\newblock
\urldef\tempurl%
\url{https://doi.org/10.1177/00222194060390060201}
\showDOI{\tempurl}


\bibitem[\protect\citeauthoryear{Werner and Scior}{Werner and Scior}{2022}]%
        {werner_scior_2022}
\bibfield{author}{\bibinfo{person}{Shirli Werner} {and}
  \bibinfo{person}{Katrina Scior}.} \bibinfo{year}{2022}\natexlab{}.
\newblock \bibinfo{booktitle}{\emph{Intellectual Disability Stigma: The State
  of the Evidence}}.
\newblock \bibinfo{publisher}{Cambridge University Press},
  \bibinfo{pages}{158–184}.
\newblock
\urldef\tempurl%
\url{https://doi.org/10.1017/9781108920995.011}
\showDOI{\tempurl}


\bibitem[\protect\citeauthoryear{Wu, Petersen, Ahmad, Burlinson, Tanis, and
  Szafir}{Wu et~al\mbox{.}}{2021}]%
        {chi}
\bibfield{author}{\bibinfo{person}{Keke Wu}, \bibinfo{person}{Emma Petersen},
  \bibinfo{person}{Tahmina Ahmad}, \bibinfo{person}{David Burlinson},
  \bibinfo{person}{Shea Tanis}, {and} \bibinfo{person}{Danielle~Albers
  Szafir}.} \bibinfo{year}{2021}\natexlab{}.
\newblock \showarticletitle{Understanding Data Accessibility for People with
  Intellectual and Developmental Disabilities}. In
  \bibinfo{booktitle}{\emph{Proceedings of the 2021 CHI Conference on Human
  Factors in Computing Systems}} \emph{(\bibinfo{series}{CHI '21})}. Article
  \bibinfo{articleno}{606}, \bibinfo{numpages}{16}~pages.
\newblock
\showISBNx{9781450380966}
\urldef\tempurl%
\url{https://doi.org/10.1145/3411764.3445743}
\showDOI{\tempurl}


\bibitem[\protect\citeauthoryear{Wu, Tanis, and Albers~Szafir}{Wu
  et~al\mbox{.}}{2019}]%
        {viscomm}
\bibfield{author}{\bibinfo{person}{Keke Wu}, \bibinfo{person}{Shea Tanis},
  {and} \bibinfo{person}{Danielle Albers~Szafir}.}
  \bibinfo{year}{2019}\natexlab{}.
\newblock \showarticletitle{Designing Communicative Visualization for People
  with Intellectual Developmental Disabilities}.
\newblock \bibinfo{journal}{\emph{IEEE VISComm}} (\bibinfo{date}{08}
  \bibinfo{year}{2019}).
\newblock
\urldef\tempurl%
\url{https://doi.org/10.31219/osf.io/zbjhr}
\showDOI{\tempurl}


\bibitem[\protect\citeauthoryear{Yang, Marriott, Butler, Goncu, and
  Holloway}{Yang et~al\mbox{.}}{2020}]%
        {yang2020tactile}
\bibfield{author}{\bibinfo{person}{Yalong Yang}, \bibinfo{person}{Kim
  Marriott}, \bibinfo{person}{Matthew Butler}, \bibinfo{person}{Cagatay Goncu},
  {and} \bibinfo{person}{Leona Holloway}.} \bibinfo{year}{2020}\natexlab{}.
\newblock \showarticletitle{Tactile Presentation of Network Data: Text, Matrix
  or Diagram?}. In \bibinfo{booktitle}{\emph{Proceedings of the 2020 CHI
  Conference on Human Factors in Computing Systems}}. \bibinfo{pages}{1--12}.
\newblock
\urldef\tempurl%
\url{https://doi.org/10.1145/3313831.3376367}
\showDOI{\tempurl}


\bibitem[\protect\citeauthoryear{Zhao, Plaisant, Shneiderman, and Lazar}{Zhao
  et~al\mbox{.}}{2008}]%
        {zhao2008data}
\bibfield{author}{\bibinfo{person}{Haixia Zhao}, \bibinfo{person}{Catherine
  Plaisant}, \bibinfo{person}{Ben Shneiderman}, {and} \bibinfo{person}{Jonathan
  Lazar}.} \bibinfo{year}{2008}\natexlab{}.
\newblock \showarticletitle{Data Sonification for Ssers with Visual Impairment:
  A Case Study with Georeferenced Data}.
\newblock \bibinfo{journal}{\emph{ACM Transactions on Computer-Human
  Interaction (TOCHI)}} \bibinfo{volume}{15}, \bibinfo{number}{1}
  (\bibinfo{year}{2008}), \bibinfo{pages}{1--28}.
\newblock
\urldef\tempurl%
\url{https://doi.org/10.1145/1352782.1352786}
\showDOI{\tempurl}


\end{thebibliography}
\end{document}